\documentstyle[sprocl]{article}

\input{psfig}
\def\be{\begin{equation}}
\def\ee{\end{equation}}
\def\bea{\begin{eqnarray}}
\def\eea{\end{eqnarray}}
\begin{document}
%                                              \hfill Fermilab-CONF-98/055-E \\

\title{SUPERSYMMETRY AT THE TEVATRON ?}

\author{ S. LAMMEL }

\address{Fermi National Accelerator Laboratory, Wilson Road,
P.O.\ Box 500, \\Batavia, IL 60510, USA \\E-mail: lammel@fnal.gov}


\maketitle\abstracts{These lectures contain an introduction to the
search for supersymmetry at hadron colliders.
The Tevatron is one of high-energy physics most sophisticated tools.
The high center-of-mass energy of its proton--antiproton collisions
makes it an ideal place to search for physics beyond the Standard
Model, such as supersymmetry.
Two experiments, CDF and D\O , completed a long data taking period
in summer of 1995, yielding over $100 \, {\rm p b}^{-1}$ of proton--%
antiproton interactions. The data recorded by the experiments are
still being analysed.
The lectures outline the strategies in the search for supersymmetry
at the Tevatron and examine the major analyses in detail. Results
obtained by the two experiments are included where available.}

\section{Introduction}
In the 1970's the Standard Model (SM)~\cite{sm} of particle physics
emerged. This theory of electromagnetic, weak, and strong interactions
has been tested by experiments during the past decade and found to be
in remarkable agreement. Although no deviations between experiment
and theoretical predictions are observed, the SM cannot be a fundamental
theory.
With the observation of the last missing quark~\cite{cdf_top,d0_top}
the matter sector of the SM is essentially complete and time for
physics beyond the Standard Model has come.

Supersymmetry (SUSY)~\cite{susy_intro}, a now more than 20 year old idea
\cite{susy_orig}, is the most popular candidate for such new physics.
This boson--fermion symmetry leads to a doubling of the particle spectrum,
as each particle now has a superpartner. Experiments should be able to
verify the existence of superpartners easily. However, with supersymmetry
being broken the new particles could be very heavy and thus outside the
sensitivity of previous/current experiments.
High-energy physics experiments at the energy frontier are continuously
probing new energy regions and could see at any time first signals from
supersymmetric particles.

The CDF~\cite{CDF_exp_home} and D\O ~\cite{D0_exp_home} experiments at
the Fermilab Tevatron are such experiments. The experiments record and
analyse proton--antiproton interactions at a center-of-mass energy of
$1.8 \, {\rm TeV}$.
Experiments at hadron colliders are very challenging and sophisticated
detectors are required to resolve the signals from individual particles
in the dense hadronic environment and to disantangle the complex events.
Because of ever remaining backgrounds, analysis of the events has to be
done on a pure statistical basis and cannot be done on an event-by-event
basis.

The search for signals of new supersymmetric particles is one of the
many physics goals of the two Tevatron experiments. In section~2 we
discuss general issues related to hadron colliders; section~3 is devoted to
possible signatures from supersymmetry and how they are identified;
in section~4 we develop our search strategies; section~5 describes
the experimental setup; in section~6 we then examine several searches
in detail; and section~7 gives a brief outlook at the future of the
Fermilab Tevatron collider.

\section{Hadron Colliders}
Supersymmetric particles may be produced in energetic collisions of
ordinary particles. There are two common approaches to create such
collisions: fixed-target and colliding beams.

In the fixed-target approach a beam of particles is accelerated
and shot onto a stationary target. The center-of-mass energy
($\sqrt{s}$) is given by
\[ s = 2 E_{\mbox{beam}} m_{\mbox{target}}, \]
where $E_{\mbox{beam}}$ is the energy of the particle beam and
$m_{\mbox{target}}$ is the mass of the target particles.
In the search for very heavy particles fixed-target experiments are
less important as the energy available for the production of new
particles rises only with the square root of the beam energy.

In the colliding beams approach two particle beams are accelerated
and brought to head on collision. The center-of-mass energy is now
proportional to the beam energy. We then have to choose the type of
particle for our beams. Since acceleration is done via electromagnetic
fields, in principle any stable (or long lived) charged particle can
be used. Electrons and protons (plus their antiparticles) are the most
commonly used particle beams in high-energy physics. The particle beams
are easy to produce and even antiparticle beams can be produced with
sufficient intensity (although, this is a bit more complicated). There
are two advantages of having particle and antiparticle beams: 1) the particles
can annihilate in the collisions, making the full center-of-mass energy
available for the production of heavy particles and 2) in a circular
machine particle and antiparticle beams can share much of the facility,
with the antiparticle beam moving on the same orbit but in opposite
direction.

In our search for new supersymmetric particles we prefer colliding
beams of particles and antiparticles. We now have the choice between
electron--positron or proton--antiproton beams. Without technical
limitations the choice would be simple. However, technical issues
associated with the accelerators, experimental issues regarding the
detectors, as well as pure physics issues, make the choice more
complicated.
Most of the issues arise from the fact that the electron is so light
and the fact that the proton is not an elementary particle.

The Large Electron Positron collider (LEP) at CERN is currently the most
energetic electron--positron collider. It is a circular machine with
a circumference of over $26 \, {\rm k m}$.
Electrically charged particles radiate when they are forced onto a
circular orbit. The electrons and positrons in LEP loose about $1.8 \,
{\rm GeV}$ per revolution through synchrotron radiation at the
highest beam energy of $96 \,{\rm GeV}$. This energy loss has to be
compensated through regular re-acceleration. The amount of energy that
one can pump back into the beam at each rotation limits the beam energy
for a given radius.

The Stanford Linear Collider (SLC) is currently the most energetic
linear electron--positron collider. The machine has a beam energy of
$50 \,{\rm GeV}$. The challenge for linear colliders is the size of
the beams at the interaction region and their overlap, which, together
with the number of particles per beam and the beam-beam collision
frequency, determines the luminosity.
The luminosity is the measure of how many collisions occur. It is
inversely proportional to the effective cross-sectional area of the
beam overlap.

Proton--antiproton colliders don't have those problems (but others).
Synchrotron radiation is proportional to $(1 / m^{4})$. With the
proton being almost $2000$ times heavier than the electron, synchrotron
radiation is lower by a factor of more than $10^{13}$. Circular machines
are thus a good approach for proton--antiproton colliders.
The Fermilab Tevatron is the world's most energetic collider. The machine
has a circumference of over $6 \, {\rm k m}$ and accelerates proton and
antiproton beams to $900 \,{\rm GeV}$.
The big problem of proton--antiproton colliders comes from the proton
not being an elementary particle like the electron. The proton is made
up of quarks and gluons. In the parton model the proton has two up
valence quarks and one down valence quark. There is also a non-zero
probablility of finding other partons, i.e.\ gluons and quarks or
antiquarks of different flavour inside the proton.
When a proton collides with an antiproton, it is normally two of these
partons that collide.
The energy of the colliding partons is only a fraction of the energy of
the proton or antiproton. It is different for the proton and antiproton
and from collision to collision. In addition we don't know the types of
the partons that collided.

\begin{figure}[tb]
\hspace{12mm}
\psfig{figure=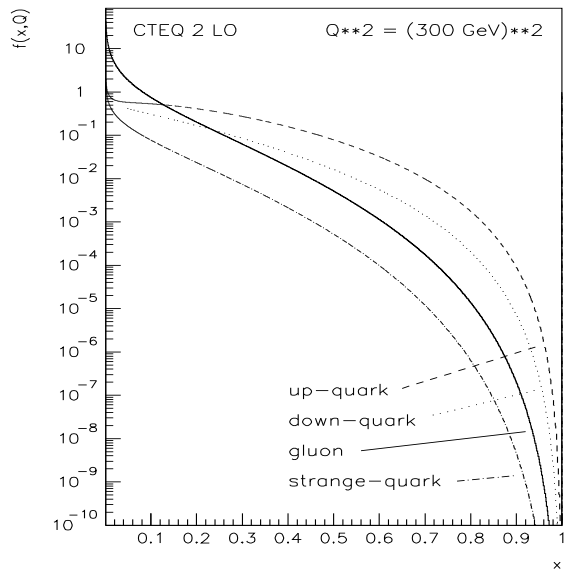,height=80mm,width=80mm}
\caption{Proton structure functions as a function of momentum fraction $x$.
\label{fig:pdf_p_300}}
\end{figure}

The probability for finding a parton of a specific type inside a proton
or antiproton is described by the structure or parton distribution
functions (PDFs). The probability depends on the momentum fraction of
the parton, $x$, and the energy transfer of the collision, $Q^{2}$.
Figure~\ref{fig:pdf_p_300} shows the proton structure functions for
several parton types as function of momentum fraction of the parton.
At low $x$ gluons dominate while at high $x$ valence quarks prevail.
The probability falls steeply, even for valence quarks, when $x$ is
larger than about $0.1$.

The proton structure functions cannot be calculated theoretically. They
describe a process outside perturbative QCD. However, if the structure
function is measured at one $Q^{2}$ it can be calculated at other values
of $Q^{2}$ and $x$. The proton structure functions are rather well known
for a large region of $x$.

\subsection{Luminosity versus Energy}
The partons that collide have only a fraction of the original proton
or antiproton momentum. The center-of-mass energy ($\sqrt{\hat{s}}$)
of the collision is given by
\[ \hat{s} \simeq x_1 x_2 s, \]
where $x_1$ ($x_2$) is the $x$ of the parton from the proton
(antiproton).
In order to increase the probability of an energetic collision we
can either increase the energy of the proton and antiproton beam or
we could increase the number of proton--antiproton collisions and thus
sample the structure functions to larger $x$ values.
Through the structure functions, energy and luminosity of a hadron
collider are connected. However, as the luminosity is increased we
are not only increasing the probability of energetic parton collisions
but also the number of less energetic parton collisions.

Particle beams are made of bunches, with a large number of particles,
about $10^{10}$ per bunch. The luminosity can then be increased by
either filling the collider with more bunches or by increasing the
number of particles in each bunch.
If more particles are filled into a bunch
the probability of having more than one proton--antiproton collision
per bunch crossing increases. Although, most of these collision will
be rather soft parton collisions, they nevertheless spread their
collision products into the experiment. Such additional interactions
not only complicate the experimental setup but also make the
reconstruction and interpretation of the observed interactions more
difficult.
Increasing the number of bunches, decreases the time between beam
crossings and thus requires faster measurements. Most detectors are
designed for a specific minimum beam crossing interval.

Increasing the beam energy is not limited by synchrotron radiation or
beam acceleration. Magnetic fields are used to keep the particles on
the circular orbit. As the beam energy increases, stronger magnetic
fields are required to hold the particles on the same orbit. The
maximum beam energy is proportional to the magnetic field and the
radius of the collider. Those magnets (or the size of the machine)
turn out to be the limiting factor for proton--antiproton colliders.
The Fermilab Tevatron has $774$ superconducting dipols with magnetic
fields of $4.4 \, \mbox{Tesla}$ to keep the protons and antiprotons
on a circular orbit of $r = 1 \, {\rm k m}$. The Large Hadron Collider
(LHC) under construction at CERN will use superconducting magnets of
about $10 \, \mbox{Tesla}$.

\subsection{Pros and Cons of Hadron Colliders}
The biggest advantage of hadron colliders is that we can build them.
We know quite well how to build a proton--antiproton collider with a
center-of-mass energy 10 time higher than the Tevatron. LHC will be close
to that energy. The required R\&D is rather small compared to building
an electron--positron collider with 10 times the center-of-mass energy of
LEP. The biggest disadvantage is that the experiments are much more
complicated. Both multiple interactions and short beam crossing interval
require challenging R\&D projects before a detector can be built and
sophisticated computer programs to analyse and interpret the observed
collisions.

If we take a look at the physics then there is another advantage of
hadron colliders: Since the colliding particles are coloured, the
production cross section for coloured objects is very large. For instance,
a pair of supersymmetric quarks (squarks) can be produced via strong
interaction. (In an electron--positron collision those would be produced
through a virtual photon or $Z$, i.e.\ electroweak interaction.)
The unknown center-of-mass energy of the parton--parton interaction is
clearly a disadvantage. Together with the complex events, this makes
individual events almost meaningless, and requires statistical analyses.

\section{Signatures of Supersymmetry}
What are possible signatures of supersymmetry and can we identify them
in proton--antiproton collisions? The two questions go hand in hand.
Signatures from supersymmetry are only good if there are no similar
signatures from Standard Model processes.

Instead of looking at various different SUSY scenarios and production
processes to see what signatures they will yield inside a Tevatron
detector, we will go over more generic signatures and discuss their
origin in both SM and SUSY. We can then see which of those signatures
occur rarely in Standard Model processes and can be used as starting
point for new particle searches.

\subsection{Missing Transverse Energy}
Since the longitudinal momentum of the colliding partons is not equal,
the parton--parton collision is not at rest in the detector but has an
unknown boost along the beam direction. The boost is of little interest
as it does not provide insights into the interaction. This makes the
plane perpendicular to the beam direction, the transverse plane, the
meaningful plane in proton--antiproton physics. Most measurements are
projected into this plane. Transverse components are indicated through
a ``${\rm T}$'' subscript.

The proton and antiproton inside the beams have no significant momentum
prependicular to the beam direction. Momentum conservation than requires
that the sum (4-momentum vector sum) of all final state particles also
has no momentum perpendicular to the beam direction.

Neutrinos are only weakly interacting. They leave no ionization trail
in the tracking detectors and deposit no energy in the calorimeters
but escape the experiment quasi-undetected. However, the 4-momentum
vector sum of all detected final state particles now has a significant
transverse momentum due to the missing neutrino in the sum. The negative
of this vector sum then corresponds to the 4-momentum vector of the
neutrino. The magnitude of its projection in the transverse plane is
called the missing transverse energy ($\not\!\!\!\:E_{\rm T}$).

Events with an energetic neutrino will have a significant missing
$E_{\rm T}$, pointing into the direction of the escaping neutrino.
Very energetic neutrinos are rare in Standard Model processes. Decays
of the intermediate vector bosons, $W$s and $Z$s, can produce energetic
neutrinos. However, their production cross section is relatively small,
about $20 \, {\rm nb}$ at the Tevatron. In addition, the majority of
$W$ and $Z$ decays don't yield neutrinos.

Standard R-parity conserving SUSY has a lightest supersymmetric
particle (LSP) that must be stable. For cosmological reasons this LSP
should have no electric or strong charge. We normally assume the
lightest neutralino ($\tilde{\chi}_{1}^{0}$) to be this LSP. The LSP is
then only weakly interacting and escapes the detector like neutrinos.
Decays of all supersymmetric particles will end into a state with the
LSP. R-parity conserving SUSY also requires supersymmetric particles
to be produced in pairs. We than have not only one but two LSPs in a
SUSY event contributing to missing transverse energy.
Missing $E_{\rm T}$ is an excellent signature for standard R-parity
conserving supersymmetry. A heavy LSP will cause large missing energy.
However, missing $E_{\rm T}$ measures the vector sum of all escaping
particles. We have no information on the energy and direction of the
individual particles or how many particles escaped undetected. With many
such particles in an event there is also the chance that some of the
particles will travel in opposite direction canceling each others
$\not\!\!\!\:E_{\rm T}$ contribution.

\subsection{Hadronic Jets}
The quarks and gluons produced in the collision carry strong charge. 
As they travel away from the interaction point, the self-interaction
between the gluons pulls the lines of the colour field together. This
string gets longer and eventually reaches a point where the stored energy
allows production of quark--antiquark pairs, i.e.\ new hadrons. Since the
transverse momentum involved in this hadron production is small, of the
order of the hadron mass, the hadrons travel along the direction of the
original quark or gluon, forming a collimated ``jet''. The process is a
result of the confinment of quarks and gluons inside hadrons and is called
fragmentation.

Jets from light quarks and gluons are indistinguishable. Jets originating
from charm or bottom quarks can contain energetic leptons from semileptonic
decays and a leading particle with on average large transverse momentum
with respect to the jet direction (due to the more massive quarks).
B-hadrons are long lived and can travel a measurable distance before they
decay. Some of the hadrons in a b-jet then have a vertex displaced from
the interaction vertex. A high resolution vertex detector can detect such
displacements and ``tag'' the jet as containing a B-hadron.

Hadronic jets are very common in proton--antiproton collisions. Almost
all inelastic interactions produce jets. The inelastic cross-section
is very large, about $60 \, {\rm mb}$. Jets, originating from charm or
bottom quarks, however, are less common than jets originating from light
quarks or gluons.

\subsection{Lepton Identification}
Of the three types of charged leptons electrons and muons are most
easily detectable. Electron detection is based on the characteristic
electromagnetic shower. It is supplemented by tracking information in
most experiments. When the electron enters the calorimeter it radiates
photons, which convert to electron--positron pairs, which radiate
new photons and produce new electron--positron pairs. The number of
particles increases exponentially with depth until the energy falls
below the threshold for electron--positron pair production.
Electromagnetic shower detectors are built from materials with high Z
and small radiation length. Lead is thus most commonly used.

Muons are not stable. However, they are so long lived that they decay
outside the detector. Muons are minimum ionizing particles. They are the
only charged particles that can traverse a large amount of material with
little energy loss. Most detectors make use of this feature and surround
the calorimeter with additional shielding that only muons can traverse
without interacting. Charged particle detectors behind this shielding
then provide a rather simple muon identification. The momentum of the
muons is measured with the help of magnetic fields by bending them from
a straight path onto a trajectory depending on their momenta.

Tau leptons are much harder to detect. Taus decay inside the detector
into an odd number of charged particles. The jet is very narrow with a
large electromagnetic component due to the presence of $\pi ^{0}$s in
most decay modes. Most experiments can detect them by statistical
methods with efficiencies up to about 50\% .

Leptons are not too common in proton--antiproton collisions. Sources for
leptons are the Drell-Yan process, leptonic decays of the intermediate
vector bosons, $W$ and $Z$, leptonic decays of vector mesons, $J/\psi $
and $\Upsilon $, and semileptonic decays of heavy quarks, charm and bottom.
The Drell-Yan production cross-section falls rapidly with dilepton mass.
It is about $20 \, {\rm pb}$ above the $\Upsilon $ resonances. Production
of vector mesons is rare and yields events with the dilepton on the mass
resonance. Leptons from semileptonic decays of charm and bottom are
inside or close to a hadronic jets. For a high-$p_{\rm T}$ lepton a hard
fragmentation of an energetic charm or bottom quark is required. The
more energetic the quark, the larger the $Q^{2}$ of the interaction has
to be, and the smaller is the cross-section.

\subsection{Photons}
Photons produce electromagnetic showers identical to electrons with the
cascade starting from a photon instead of an electron. The difference to
the electron signature is the ``missing'' charged particle track.

Photons have become famous in supersymmetry searches about two years ago 
due to an event with photons and $\not\!\!\!\:E_{\rm T}$ observed by the
CDF experiment. Production of energetic photons from Standard Model
processes is rare
\cite{CDF_photon}, making photons a good signature of interesting physics.
Energetic photons can come from direct production and hard bremsstrahlung.

The problem with photons comes from the large number of jets that are so
common in proton--antiproton collisions. If a jet has a very hard
fragmentation and one $\pi ^{0}$ takes most of the energy, the jet could
mimic a photon signature in the calorimeter. Although the probability of
such a fluctuation is tiny, the number of jets is huge.

\section{Search Strategies}
Let's take a look at the particle spectrum of the MSSM
(Table~\ref{tab:mssm_particles}) and see which of the superpartners could
be accessible to us at a hadron collider.

\begin{table}[tbp]
\begin{center}
\caption{Particle Contents of the Minimal Supersymmetric extension of
the Standard Model,
\label{tab:mssm_particles}}
\begin{tabular}{|l|l|l|}
                                                                    \hline
\multicolumn{1}{|c|}{\bf spin = $0$} &
\multicolumn{1}{|c|}{\bf spin = $1/2$} &
\multicolumn{1}{|c|}{\bf spin = $1$}                             \\ \hline
                                                                    \hline
sleptons:
\( \left( \begin{array}{c} \tilde{e} \\
                          \tilde{\nu _{e}} \end{array} \right) \)
\( \left( \begin{array}{c} \tilde{\mu} \\
                        \tilde{\nu _{\mu}} \end{array} \right) \)
\( \left( \begin{array}{c} \tilde{\tau} \\
                       \tilde{\nu _{\tau}} \end{array} \right) \)
&
leptons:
\( \left( \begin{array}{c} e \\ \nu _{e} \end{array} \right) \)
\( \left( \begin{array}{c} \mu \\ \nu _{\mu} \end{array} \right) \)
\( \left( \begin{array}{c} \tau \\ \nu _{\tau} \end{array} \right) \)
&                                                                \\ \hline
squarks:
\( \left( \begin{array}{c} \tilde{u} \\
                              \tilde{d} \end{array} \right) \)
\( \left( \begin{array}{c} \tilde{c} \\
                              \tilde{s} \end{array} \right) \)
\( \left( \begin{array}{c} \tilde{t} \\
                              \tilde{b} \end{array} \right) \)
&
quarks:
\( \left( \begin{array}{c} u \\ d \end{array} \right) \)
\( \left( \begin{array}{c} c \\ s \end{array} \right) \)
\( \left( \begin{array}{c} t \\ b \end{array} \right) \)
&                                                                \\ \hline
                                                                    \hline
& gluinos & gluons                                               \\ \hline
\( \begin{array}{l} \\ \mbox{charged Higgs: } H^{\pm} \end{array} \)
&
charginos:
\( \left.  \begin{array}{l} \tilde{\chi}^{\pm}_{1} \\
                       \tilde{\chi}^{\pm}_{2} \end{array} \right\}
   \left\{ \begin{array}{l} \tilde{W}^{\pm} \\
                              \tilde{H}^{\pm} \end{array} \right.  \)
&
\( \begin{array}{l} W^{\pm} \mbox{boson} \\ \ \end{array} \)  \\ \hline
\( \begin{array}{l} \\ \\ \\
    \mbox{Higgses: } h^{0}, A^{0}, H^{0} \end{array} \)
&
neutralinos:
\( \left.  \begin{array}{l} \tilde{\chi}^{0}_{1} \\ \tilde{\chi}^{0}_{2} \\
      \tilde{\chi}^{0}_{3} \\ \tilde{\chi}^{0}_{4} \end{array} \right\}
   \left\{ \begin{array}{l} \tilde{\gamma} \\ \tilde{Z}^{0} \\
      \tilde{H}^{0}_{1} \\ \tilde{H}^{0}_{2} \end{array} \right.  \)
&
\( \begin{array}{l} \gamma \\ Z^{0} \\ \ \\ \  \end{array} \)      \\ \hline
\end{tabular}
\end{center}
\end{table}

The most interesting superpartners for searches at hadron colliders are
the coloured particles, i.e.\ the squarks and gluons. Their production
cross-section is rather large~\cite{DESY_nlo}, for $200 \, {\rm GeV}/c^2$
squarks and
gluinos the production cross-section is about $\sigma _{\tilde{g}\tilde{g}}
= 3 \, {\rm pb}$, $\sigma _{\tilde{g}\tilde{q}} = 13 \, {\rm pb}$,
$\sigma _{\tilde{q}\tilde{\bar{q}}} = 12 \, {\rm pb}$, and
$\sigma _{\tilde{q}\tilde{q}} = 2 \, {\rm pb}$.
The production cross-section falls quickly as the mass of the particles
increases. For $250 \, {\rm GeV}/c^2$ squarks and gluinos we expect
$\sigma _{\tilde{g}\tilde{g}} = 0.4 \, {\rm pb}$,
$\sigma _{\tilde{g}\tilde{q}} = 1.9 \, {\rm pb}$,
$\sigma _{\tilde{q}\tilde{\bar{q}}} = 2.7 \, {\rm pb}$, and 
$\sigma _{\tilde{q}\tilde{q}} = 0.3 \, {\rm pb}$, i.e.\ only about 300
gluino--gluino events for an integrated luminosity of $100 \, {\rm pb}^{-1}$.

\subsection{Direct versus Cascade Decays}
Let's assume we would have actually produced a couple of gluinos and
squarks at the Tevatron. They are not stable but will decay into the
lightest supersymmetric particle, LSP, the $\tilde{\chi}^{0}_{1}$ in our
model.

In the region of relatively small gluino and squark masses we have
direct decays into the LSP, which is then mostly photino (ignoring the
region of small Higgsino parameter for now). As we go to higher gluino
and squark masses, decays into charginos and heavier neutralinos are
kinematically allowed. Especially decays into the chargino can become
the dominant decay. The charginos and heavier neutralinos are not stable
and decay into quarks or leptons and the LSP. Those cascade decays of
gluinos and squarks are important in the mass range currently explored
by experiments~\cite{Baer_cascade}.

\begin{figure}[tbp]
\begin{center}
\begin{picture}(252,252)
\put(2,180){\Large \bf $m_{\tilde{q}}$}
\put(180,2){\Large \bf $m_{\tilde{g}}$}
\put(25,25){\vector(0,1){220}}
\put(25,25){\vector(1,0){220}}
\put(30,220){\large $\tilde{g} \rightarrow q \overline{q}' \tilde{\chi}^{\pm}$}
\put(71,208){\large $\hookrightarrow q \overline{q}' \tilde{\chi}^{0}_{1}$}
\put(39,193){\large $\rightarrow q \overline{q} \tilde{\chi}^{0}_{2}$}
\put(68,181){\large $\hookrightarrow q \overline{q} \tilde{\chi}^{0}_{1}$}
\put(30,92){\large $\tilde{q} \rightarrow q \tilde{g}$}
\put(30,75){\large $\tilde{g} \rightarrow q \overline{q} \tilde{\gamma}$}
\put(25,30){\line(5,4){210}}
\put(70,49){\large $\tilde{g} \rightarrow q \overline{\tilde{q}}$}
\put(70,32){\large $\tilde{q} \rightarrow q \tilde{\gamma}$}
\put(175,64){\large $\tilde{q}_{\rm \, L} \rightarrow q' \tilde{\chi}^{\pm}$}
\put(175,47){\large $\tilde{q}_{\rm \, L,R} \rightarrow q \tilde{\chi}^{0}_{2}$}
\end{picture}
\end{center}
\caption{Gluino--squark mass plane with dominant decay modes.
\label{fig:direct_cascade}}
\end{figure}
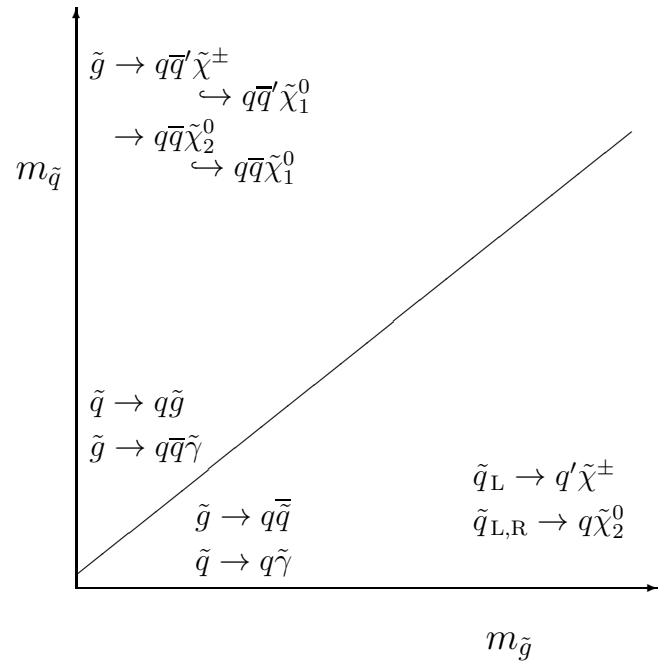

The signature of, for instance, gluino--gluino production changes as
we go from a region of light gluinos and direct decays to heavier
gluinos cascading into the LSP. The missing $E_{\rm T}$ plus 4 jets
signature now, e.g., becomes a missing $E_{\rm T}$ plus 8 jets signature.
While the $\not\!\!\!\:E_{\rm T}$ was the dominant part of the signature
at small gluino and squark masses, the jets are becoming a more important
part of the signature for larger gluino and squark masses.

The missing $E_{\rm T}$ based search is the classic SUSY search
strategy at hadron colliders. It was proposed first for Fermilab fixed
target experiments~\cite{FNAL_missing_et}, later used at the
$Sp\overline{p}S $at CERN~\cite{SPPS_missing_et}, and is used now at
the Tevatron~\cite{D0_missing_et,CDF_missing_et}.

\subsection{Weak versus Strong Production}
\begin{figure}[tbp]
\hspace{8mm}
\psfig{figure=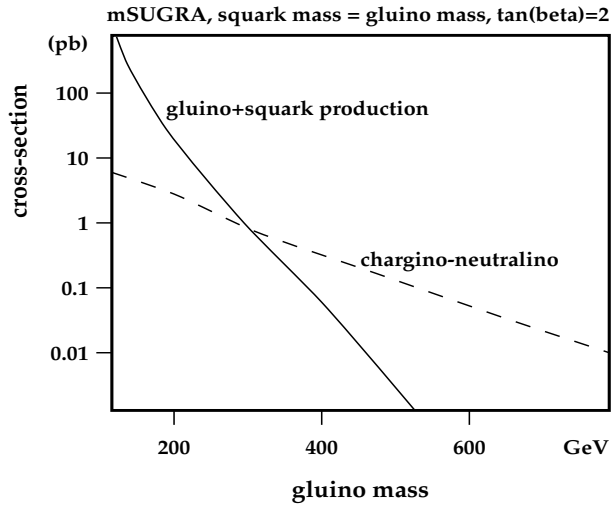,height=74mm,width=90mm}
\caption{Production cross-sections for chargino--neutralino and
gluino/squark pair production for a specific, minimal SUGRA, SUSY scenario.
\label{fig:weak_strong}}
\end{figure}

The production cross-section of coloured superpartners is very large at
proton--antiproton colliders. However, it falls steeply as the masses of
the superpartners increases. Weak production cross-sections are rather
small, e.g.\ the production of chargino--neutralino pairs. It involves
the valence quarks in the proton and antiproton, who's presence is less
probable than finding gluons but who's $x$ spectrum is much harder.
The weak production cross-sections then fall more slowly as
the superpartner mass increases. As for large $x$ values valence quarks
prevail, there is a point in SUSY parameter space at which weak production
cross-sections of, for example, chargino--neutralino pairs surpass the
strong cross-sections of, for example, gluino--gluino pairs
(Figure~\ref{fig:weak_strong}).

Searches based on weak production processes have become an important
supplement in the search for supersymmetry at the Tevatron. In addition
to the harder cross-section the simpler final states (due to the small
or absent colour flow) of such processes has made searches based on them
very attractive. The trilepton based search for chargino--neutralino
production~\cite{CDF_trilepton,D0_trilepton} is an example of such an
analysis.

\section{Collider and Experiments}
The Fermi National Accelerator Laboratory is located about 35 miles west
of Chicago. The heart of Fermilab is the Tevatron, a superconducting
proton--antiproton synchrotron and storage ring with a circumference of
$6.28 \, {\rm km}$.

\subsection{The Fermilab Tevatron}
\begin{figure}[tbp]
\psfig{figure=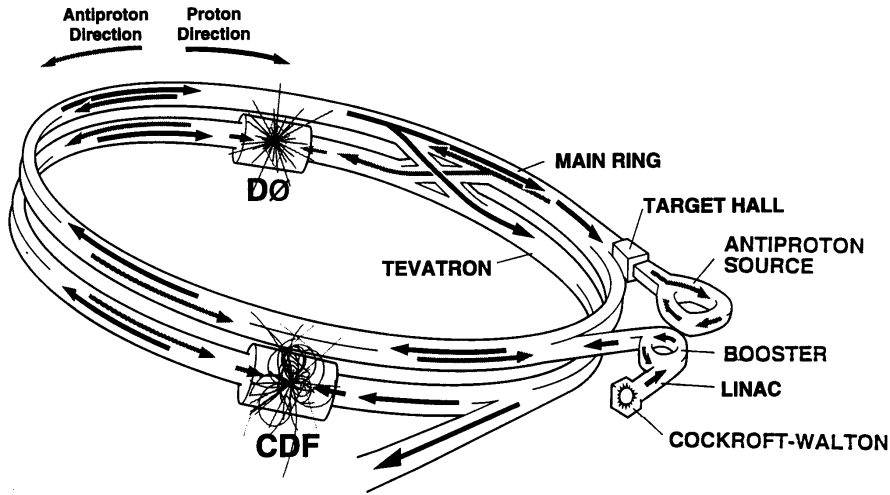,height=71mm}
\caption{A schematic layout of the Tevatron and its associated accelerators.
\label{accelerator_complex}}
\end{figure}

A schematic layout of the Tevatron and its associated accelerators is
shown in Figure~\ref{accelerator_complex}. The acceleration process starts
with $H^-$ ions in a Cockroft-Walton generator. The electrons are stripped
from the hydrogen ions and the remaining protons accelerated in multiple
steps. After acceleration in a LINAC and booster bunches of protons are
inserted
into the Main Ring where they are accelerated to $150 \, {\rm GeV}$. The
protons can then be either transfered into the Tevatron or sent onto a
nickel target for antiproton production. The antiproton yield is rather
low, about $10^5$ protons are used to produce and collect one antiproton.
The antiprotons are focused, cooled, and stored in a pair of storage rings.
After about 24 hours enough antiprotons are collected. They are bunched and
accelerated in the Main Ring before being transfered into the Tevatron.
The Tevatron takes six proton and six antiproton bunches and accelerates
them to their final energy of $900 \, {\rm GeV}$. Proton and antiproton
beams are brought to collision in two high luminosity interaction regions,
B0 and D0. A proton bunch collides with an antiproton buch $286,000$
times per second or every $3.5 \, {\rm \mu sec}$.

During the 1992/93 collider running period the Tevatron produced about
$30 \, {\rm pb}^{-1}$ of $p \overline{p}$ interactions at each of the
high luminosity regions and an integrated luminosity of about $156 \,
{\rm pb}^{-1}$ during the 1994/95 running period.

Two experiments recorded and analysed these interactions, the Collider
Detector at Fermilab (CDF) and the D\O\ experiment. Although the detailed
design
of the detectors is very different, the basic structure is pretty much
the same for most collider detectors: tracking detectors in the innermost
region; electromagnetic and hadronic calorimeters surrounding the tracking
systems; and muon detectors on the outside.

\subsection{The CDF Detector}
\begin{figure}[tb]
\psfig{figure=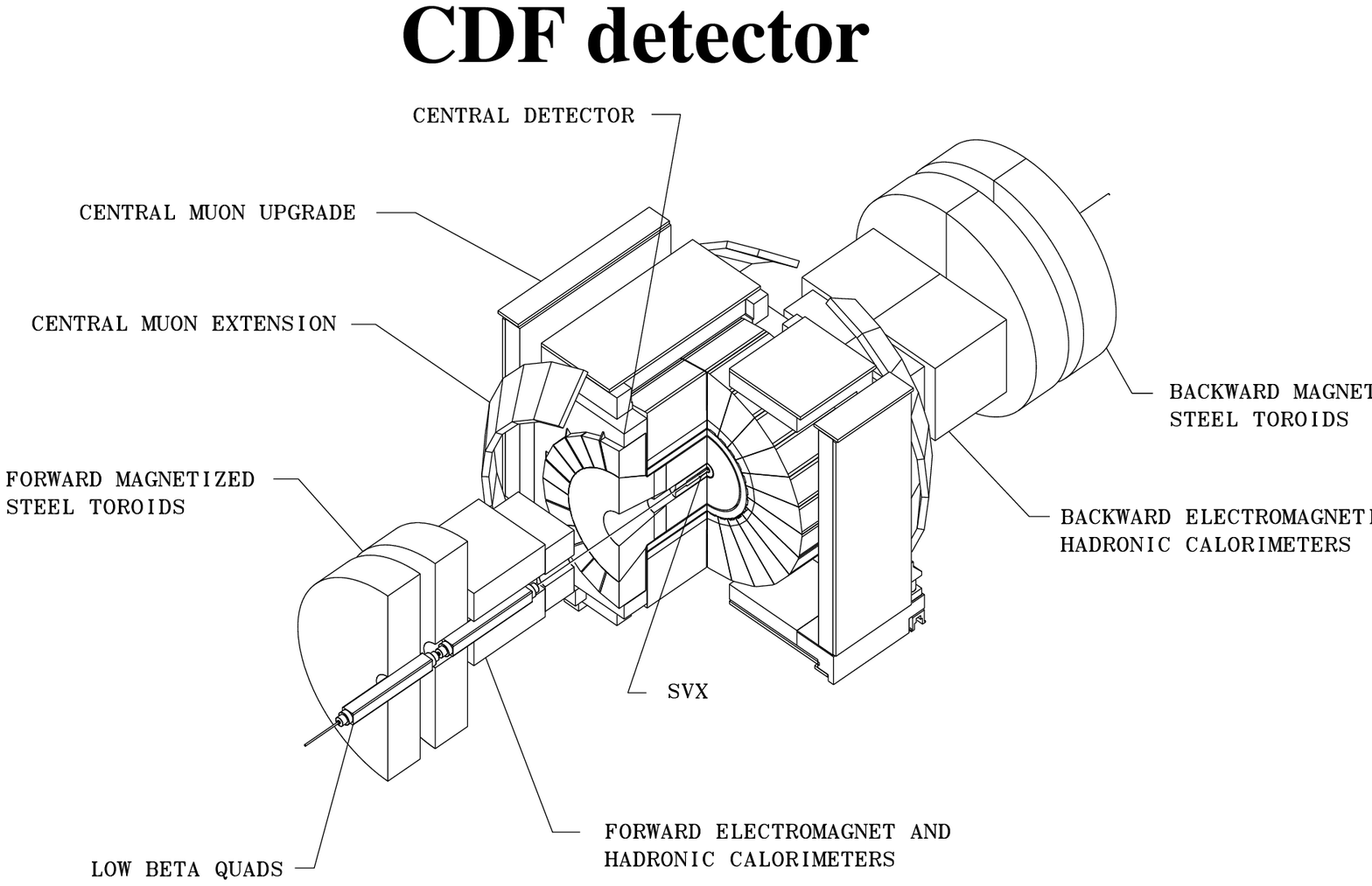,height=81mm}
\caption{Cross-section view of the CDF detector.
\label{fig:cdf_detector}}
\end{figure}

The CDF detector~\cite{CDF_detector} is a general purpose collider detector
surrounding the B0 interaction region. A cross-section of the 1994/95
configuration is shown in Figure~\ref{fig:cdf_detector}.
In the design of the detector emphasis was put on charged particle tracking.
The three tracking components are inside a strong magnetic field of $1.4 \,
\mbox{Tesla}$ that is provided by a superconducting solenoidal magnet.
The central tracking chamber (CTC) measures the curvature of charged
particles within pseudorapidity~\cite{coordinates} $| \eta | < 1.5$ and
provides a momentum resolution of
$\delta p_{\rm T}/p^{2}_{\rm T} = 0.001$. This allows precise momentum
measurements up to large transverse momenta, which is particularly important
for energetic muons.
A system of time projection chambers (VTX) provides precise $\rm r - z$
information for the charged tracks.
The innermost part of the detector is occupied by a four-layer silicon
micro-vertex detector (SVX). The SVX provides spacial measurements in the
$\rm r - \varphi $ plane and an impact parameter resolution of $( 13 + 40 /
p_{\rm T} ) \, {\rm \mu m}$ where $p_{\rm T}$ is the transverse momentum
of the track in ${\rm GeV}/c$. With the help of the SVX secondary vertices,
from the decay of long-lived particles, can be identified.

Electromagnetic and hadronic calorimeters cover the central region $| \eta |
< 1.1$, end-cap regions $1.1 < | \eta | < 2.4$, and forward regions $2.2 <
| \eta | < 4.2$. The central calorimeters consist of 48 wedge shaped modules,
each covering $14.5^{\circ}$ in $\varphi$ and about $1.1$ units in $\eta$.
The central electromagnetic calorimeter measures the energy of electrons
with a resolution $(\sigma _{E}/E)^{2} = (13.5\% )^{2} / E + (1\% )^{2}$.
The depth of the calorimeter at $| \eta | = 0$ corresponds to $5.3$ nuclear
interaction lengths. The hadronic calorimeter has an additional 48 modules
in the region $0.7 < | \eta | < 1.3$. The end-cap and forward electromagnetic
(hadronic) calorimeters are made out of proportional tube arrays sandwiched
with lead (steel) absorbers. The large number of calorimeter components in
the detector bring with them an even larger number of transition regions
between them. This has a significant impact in the measurement of an event's
energy imbalance.

In the central region three muon systems make the outer cover. There are
48 central muon chamber (CMU) modules in the region $| \eta | < 0.6$, backed
up by a new system of drift chambers (CMP) behind additonal steel absorber.
A third system (CMX) extends muon coverage into the region $0.6 < | \eta |
< 1.0$ but has only a $60\%$ $\varphi$ coverage.

\subsection{The D\O\ detector}
\begin{figure}[tbp]
\psfig{figure=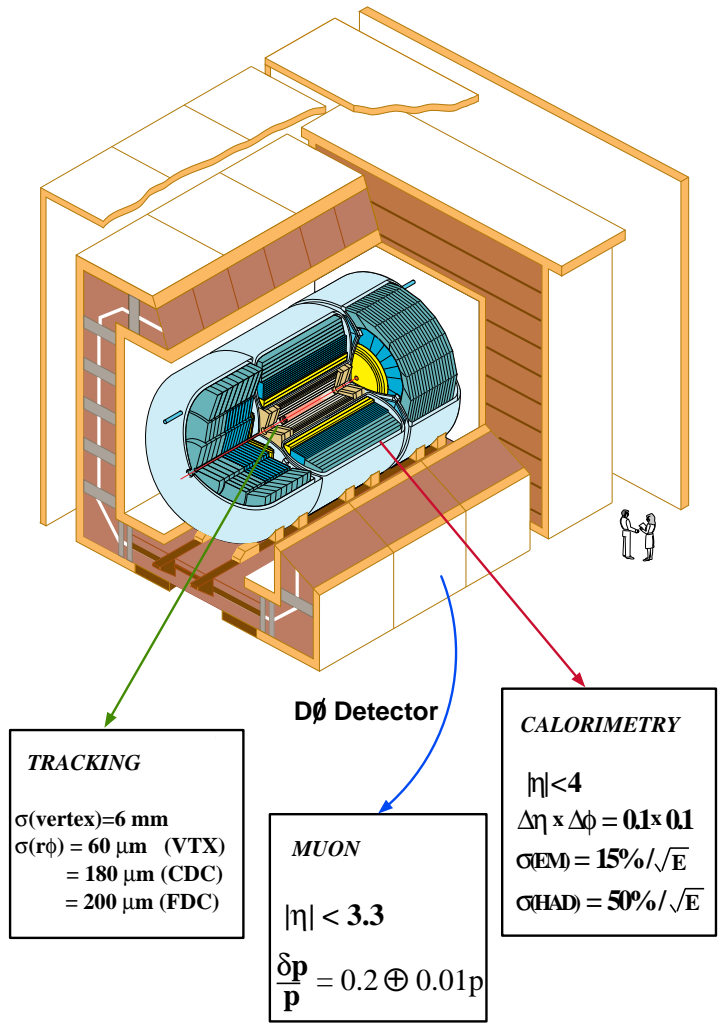,height=140mm}
\caption{View of the D\O\ detector.
\label{fig:d0_detector}}
\end{figure}

The D\O\ detector~\cite{D0_detector} is located at the second high
luminosity interaction region, D0. The design of the D\O\ detector
was optimized to have a hermetic, finely segmented, thick calorimetry
and hermetic muon detection up to large rapidities.
Figure~\ref{fig:d0_detector} shows a view of the detector.

A very compact system of tracking and transition radiation detectors occupies
the innermost region. A vertex drift chamber (VTX)  surrounds the interaction
region. It is located just outside the beryllium beam pipe and provides
$\rm r - \varphi $ information with a resolution of about $50 \, \mu {\rm m}$
for the vertex reconstruction.
Outside the VTX a transition radiation detector (TDR) provides
calorimeter-independent electron identification. The detector consists of
three units with each 393 polypropylene foils as radiator and proportional
wire chambers for X-ray detection.
Central and forward drift chambers complete the D\O\ tracking system. The
central drift chamber (CDC) extends to a radius of $74.5 \, {\rm cm}$. It
has four layers of 32 cells. Seven sense wires per cell measure $\rm r -
\varphi $ and two delay lines provide a $2 \, {\rm mm}$ $\rm z$ resolution
for the tracks of charged particles.
The forward drift chambers (FDCs) are disk shaped, extending up to $\rm z$
of $\pm 135 \, {\rm cm}$. The chambers provide $\varphi - \theta $
information for charged particle tracking down to $\theta = 5^{\circ}$.

Finely segmented uranium--liquid argon calorimeters cover the region
$| \eta | < 4$. The calorimeters are housed in three cryostats, one for
the central region $| \eta | < 1$ (CC) and one for each endcap region (EC).
Each calorimeter consists of an electromagnetic section with four depth
segments, a fine-hadronic section with three or four depth segments, and a
coarse-hadronic section with one or three depth segments. The transverse
segmentation is $\Delta \varphi = \Delta \eta = 0.1$. The energy resolution
for electrons is $(\sigma _{E}/E)^{2} = (15.7\% )^{2} / E + (0.3\% )^{2}$
and $(\sigma _{E}/E)^{2} = (41\% )^{2} / E + (3.2\% )^{2}$ for hadrons.
The cryostats are relatively massive vessels. In the transition region
from central to encap calorimeter this causes a rather large amount of
uninstrumented material. To correct for energy deposited in this material,
two arrays of scintillator counters (ICD) are mounted between CC and EC.

The Main Ring and Tevatron are both in the same tunnel, with the Main Ring
about $70 \, {\rm cm}$ above the Tevatron. At the B0 interaction region
the Main Ring
is lifted to pass above the CDF experiment. However, at the D0 interaction
region it is lifted only to about 2 meters above the Tevatron. The top
part of the coarse-hadronic sections of both CC and EC calorimeters have a
small by-pass for the Main Ring. During operation of the Main Ring this can
cause unwanted energy depositions in those calorimeters.

Three superlayers of proportional drift tubes surround the calorimeters for
muon detection. Iron toroid magnets with a field of about $2 \, \mbox{Tesla}$
between the first and second superlayer allow measurement of the muon
momentum with a resolution of $\delta p/p = 0.2 + 0.01 * p$. The central
toroid (CF) covers the region $| \eta | < 1$, the end toroids (EF) cover
$1 < | \eta | < 2.5$, and the small-angle muon system (SAMUS) extend the
coverage up to $| \eta | < 3.6$. Calorimeters and toroids provide over
$12$ nuclear interaction length and thus a natural momentum cut of $3.5 \,
{\rm GeV}/c$ for muons at $| \eta | = 0$.

\section{Specific Searches}
In the previous sections we discussed general aspects of SUSY searches
at the Tevatron. We took a look at the different signatures particles
leave in the detector, discussed which of the superpartners could be
in reach of the Tevatron, and detection capabilities of the experiments.
We now put all of this information together and look at a variety of
specific SUSY searches that have been performed or are in progress.

\subsection{Gluinos and Squarks}
Gluinos and squarks are most important in the search for supersymmetry
at the Tevatron. The reach of the experiments is very high, up to
gluino and squark masses of hundreds of ${\rm GeV}/c^{2}$. The searches
are complementary to $e^{+}e^{-}$ experiments where, for instance, the
gluino cannot be produced directly.

\subsubsection{Classic Missing $\bf E_{\rm T}$}
Let us take the D\O\ Run Ia analysis~\cite{D0_missing_et} as example.
We are interested in a
signature of transverse energy imbalance, as a possible result of stable,
only weakly interacting LSPs, and multiple jets. If we take a look back
at figure~\ref{fig:direct_cascade} we see that we expect more jets in the
case of gluino pair production and less jets in the case of dominant
squark pair production. Gluino pair production is dominant if the squarks
are heavier, squark pair production when gluinos are heavier. For
equal gluino and squark masses gluino--squark production is non-negligable.
Minimal SUGRA~\cite{sugra} models predict squarks to be heavier than gluinos.

When the Tevatron is in operation there are $286,000$ beam--beam
crossings per second. Any of those could produce gluinos and squarks.
We cannot record all of those interactions. Instead, we look very very
quickly (within $3.5 \, {\rm \mu sec}$) at parts of the event/detector
and decide if this interaction was interesting or not.
This online selection, i.e.\ triggering, is done through several levels,
allowing each successive level to take a bit more time and examine the
event more carefully.
The level 1 trigger of the experiments selected about $2 \, {\rm k Hz}$
of events from the $286 \, {\rm k Hz}$ input. For our gluino and squark
search we are interested in events with missing $E_{\rm T}$ and jets.
The level 1 trigger made a quick check if there was a tower in the
calorimeter with significant energy. An event with jets or significant
$\not\!\!\!\:E_{\rm T}$ would deposite energy in a lot of towers of the
calorimeter. For D\O\ Run Ia this energy threshold was $3 \, {\rm GeV}$
for a single tower or $5 \, {\rm GeV}$ for a three tower sum.
The level 2 trigger of the experiments rejected about 99.9\% of the events
passed by level 1 after a more carefull study. The electronics now has
several milliseconds to analyse the event. It is possible to sum the
energy in neighbouring towers of the calorimeter for a rough jet
reconstruction and calculate the transverse energy imbalance of the
calorimeter. Of interest to our missing $E_{\rm T}$ analysis are two
of the D\O\ Run Ia triggers: events with missing $E_{\rm T}$ of $35 \,
{\rm GeV}$ or more and events with three or more jets with $E_{\rm T}
> 20 \, {\rm GeV}$ and $\not\!\!\!\:E_{\rm T} > 25 \, {\rm GeV}$.
Of the $25 \, {\rm Hz}$ of events accepted by level 2 only about
$5 \, {\rm Hz}$ passed the selection criteria in level 3 and were
recorded on magnetic tape. The level 3 triggers are software implemented
with fast versions of the offline reconstruction algorithms and preliminary
calibration constants. In the D\O\ missing $E_{\rm T}$ case the jet and
$\not\!\!\!\:E_{\rm T}$ values are recalculated more accurately and the
level 2 cuts are reapplied.

The D\O\ missing $E_{\rm T}$ analysis has a three jet and a four jet path
to best address the two cases of lighter squarks and lighter gluinos. To
get the best missing $E_{\rm T}$ resolution D\O\ restricted themselves
to only single interaction events: one vertex within $\pm 70 \, {\rm cm}$
of the center of the detector.
This reduces the sample by over 40\%, from an integrated luminosity of
$13.4 \, {\rm pb}^{-1}$ to $7.4 \, {\rm pb}^{-1}$.

The three jet path addresses the region of lighter squarks: events are
required to have at least three jets with $E_{\rm T} > 25 \, {\rm GeV}$
and a $\not\!\!\!\:E_{\rm T} > 75 \, {\rm GeV}$. The four jet path is
optimized for the case of lighter gluinos. The energy of the gluinos is
now distributed among more particles, resulting in softer particles, and
thus in less energetic jets and lower missing $E_{\rm T}$. In this path
events must have at least four jets with $E_{\rm T} > 20 \, {\rm GeV}$
and a $\not\!\!\!\:E_{\rm T} > 65 \, {\rm GeV}$. The choice between fewer
but more energetic objects and more but less energetic objects is driven
by signal sensitivity on one side and background rejection on the other
side. An analysis path with three $E_{\rm T} > 20 \, {\rm GeV}$ jets and
$\not\!\!\!\:E_{\rm T} > 65 \, {\rm GeV}$ would have less sensitivity as
it is more background contaminated.

The detectors are not perfect. There are uninstrumented regions due to
support structures, signal cables from the inner detectors, supply lines,
to name just a few and transition regions between components, e.g.\ from
the central calorimeters to the end-cap calorimeters to the forward
calorimeters. In such regions, particles can be absorbed without yielding
a detectable signal and thus cause an overall energy imbalance in the
event. The chance of ``loosing'' all the energy of a particle in the
detector is rather small.
However, QCD processes produce so many events with gluons or light
quarks that there is a fair chance for some of those jets to loose a
lot of energy and for the event to have significant missing $E_{\rm T}$.
For those events the $\not\!\!\!\:E_{\rm T}$ and mismeasured jets have a
strong correlation. D\O\ found the following cuts to be very effective
in reducing mismeasured multijet events from QCD processes:
$ 6^{\circ} < \Delta \varphi ( \not\!\!\!\:E_{\rm T}, \mbox{jet} ) <
  (180 - 6)^{\circ}$ and
$ \sqrt{( \Delta \varphi ( \not\!\!\!\:E_{\rm T}, \mbox{jet}_{1} )
                                                  - 180^{\circ})^{2}
      + ( \Delta \varphi ( \not\!\!\!\:E_{\rm T}, \mbox{jet}_{2} ) )^{2} }
 < 29^{\circ}$.

Events with genuine missing $E_{\rm T}$ from Standard Model processes
are still remaining in the sample. $W$ plus multijet production is the
most significant contribution. These events have an additional energetic
lepton from the leptonic $W$ decay. Thus, the analysis rejects all events
with an
indentified electron with $p_{\rm T} > 20 \, {\rm GeV}$ or muon with
$p_{\rm T} > 15 \, {\rm GeV}$.

The remaining events were examined by eye. Jets in 8 events were found
to be caused by detector noise and removed. One event was a clear case
of a cosmic ray and removed. Two events showed a problem with the vertex
reconstruction. When it was corrected the missing $E_{\rm T}$ fell below
the cuts. $14$ events passed the three jet path and $5$ events the four
jet path.

Before we can answer the question if these events are due to gluinos and
squarks and a first sign of supersymmetry we have to calculate how many
events with the above characteristic we would expect from known Standard
Model processes and technical background sources. The missing $E_{\rm T}$
signal is tricky: any problem and technical background will most likely
result in an energy imbalance (problems are never $\varphi$-symmetric).
Sources of technical background are read-out problems of the detector
(each detector has about $150,000$ channels), accelerator beam loss (the
Main Ring is in the same tunnel just above the Tevatron and accelerating
protons for antiproton production during most of the Tevatron operation),
cosmic-ray bremsstrahlung, and detector noise to name the most important
ones.

Events from Standard Model processes that remain in the sample are
calculated with the help of Monte Carlo programs. The generated events
are passed through a detector simulation and then analysed as the real
data. Table~\ref{tab:met_bckgnd} shows the estimate of the remaining
background for the three and four jet missing $E_{\rm T}$ analyses.

\begin{table}[tbp]
\begin{center}
\caption{Expected Standard Model contribution in the D\O\ Run Ia
missing $E_{\rm T}$ analyses.
\label{tab:met_bckgnd}}
\begin{tabular}{|l|c|c|}
                                                                   \hline
 process                 & $3$ jet analysis & $4$ jet analysis  \\ \hline
                                                                   \hline
 $W \longrightarrow e \nu $    & 2.7            & 1.5 events    \\ \hline
 $W \longrightarrow \mu \nu $  & 4.0            & 1.8 events    \\ \hline
 $W \longrightarrow \tau \nu $ & 3.4            & 0.9 events    \\ \hline
 $Z \longrightarrow \nu \nu $  & 3.3            & 0.9 events    \\ \hline
 $Z \longrightarrow \mbox{other}$ & 0.9         & 0.1 events    \\ \hline
                                                                   \hline
 total:                        & $14.2 \pm 4.4$ & $5.2 \pm 2.2$ \\ \hline
\end{tabular}
\end{center}
\end{table}

The remaining contamination from QCD associated multijet production after
the $\Delta \varphi$ cuts was found to be small: $0.42$ events for the
three jet analysis and $1.6$ events for the four jet analysis.
The expected events from Standard Model processes remaining in the sample
account for all observed events.
$W$ and $Z$ plus multijet production with leptonic decays is the dominant
background contribution. The signature is similar to the SUSY signature,
the difference being the additional lepton. The lepton coverage of the
detectors is very hermetic but has holes and is limited in $\eta $. For
the $W$ and $Z$ events that remain in the sample the experiment did not
identify the lepton. The $Z \longrightarrow \nu \nu $ contribution is
small but irreducible as it matches our SUSY signature.

To set limits on gluino and squark production we have to estimate the
sensitivity of the experiment. Specific SUSY models are used for this
step. The most general R-parity conserving MSSM has 63 parameters. To
reduce the parameter space, gaugino masses are related to gauge couplings
as in supergravity grand unified theories. A common sfermion mass at the
GUT scale is used. Five degenerate squarks are assumed (any contribution
from stop is ignored). We then have as free parameters the gluino mass,
the common squark mass, $\tan (\beta)$, the Higgsino mass parameter $\mu $,
and the Higgs mass $m_{A}$.

\begin{figure}[tbp]
\psfig{figure=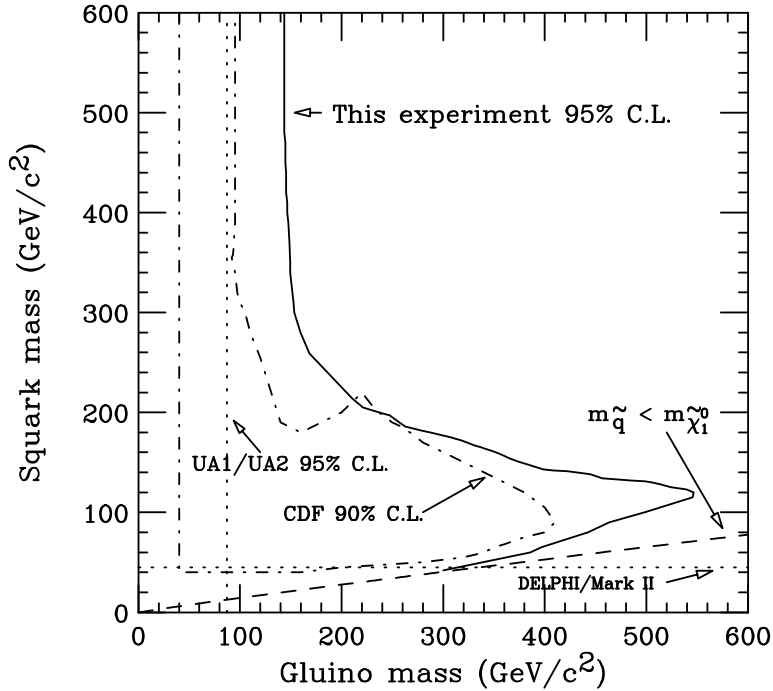,height=100mm}
\caption{Gluino and squark 95\% confidence level mass limits of the
D\O\ missing $E_{\rm T}$ plus three and four jets analyses.
\label{fig:d0_met_limit}}
\end{figure}

Figure~\ref{fig:d0_met_limit} shows the limits obtained by the D\O\
analyses. The limits are shown in the gluino--squark mass plane as those
are the most sensitive parameters. The limits vary only little for large
regions of $\tan (\beta)$ and $\mu $.

\subsubsection{Like-sign Dilepton}
The like-sign dilepton search targets gluino pair production. It makes
use of the gluino cascade decays into charginos with subsequent leptonic
chargino decays. The gluino is a Majorana particle, thus the electric
charge of the chargino in the gluino decay can be either positive or
negative: $\tilde{g} \longrightarrow u \overline{d} \tilde{\chi}^{-}$ or
$\tilde{g} \longrightarrow \overline{u} d \tilde{\chi}^{+}$.
In the case of gluino pair production we can then have a final state with
four jets, two like-sign leptons, and missing $E_{\rm T}$ from the two
neutrinos and LSPs. Leptons are not very common in proton--antiproton
interactions. Dilepton events are very rare. Like-sign dilepton events
are even more rare and together with missing $E_{\rm T}$ and jets a
striking signature.

Standard Model processes yielding like-sign dileptons are $t \overline{t}$
production with one lepton from the W decay and one from the b decay of
the other top quark, $B^{0} \overline{B^{0}}$ oscillation, and
$b \overline{b}$ production with one first and one second generation
semileptonic decay.

Let's take a closer look at the CDF Run Ib analysis. CDF selects events
with two isolated leptons, electrons or muons. The transverse momentum
requirement for the first lepton is $11 \, {\rm GeV}/c^{2}$. The lepton
is also required to be in the central region of the detector. The first
lepton then guarantees a high efficiency for the event to pass the single
electron or muon trigger. The second lepton has a lower momentum
threshold, $p_{\rm T} > 5 \, {\rm GeV}/c^{2}$ and is not restricted to
the central region. The leptons have to be of same electric charge and
both are required to be isolated in the calorimeter, $I =
\sum_{\Delta R < 0.4} \left| E_{\rm T} \right| < 4 \, {\rm GeV}$.
In our SUSY scenario the leptons come from the decay of a chargino,
$\tilde{\chi}^{\pm} \longrightarrow l^{\pm} \nu \mbox{LSP}$ and are isolated
from hadronic activity. However, like-sign dileptons from Standard Model
processes have at least one lepton from a semileptonic $b$ or $c$ decay
and are thus surrounded by hadronic activity.
The analysis requires two jets of at least $E_{\rm T} > 15 \, {\rm GeV}$
and a missing $E_{\rm T}$ of at least $25 \, {\rm GeV}$. The value of
these cuts is much lower than what was used in the missing $E_{\rm T}$
plus multijet search. The reason is that the leptons are a very good
signature and we can now afford lower thresholds without large background
contamination. The $\not\!\!\!\:E_{\rm T}$ is now the result of four
particles escaping detection, two neutrinos and two LSPs. Some of those
will balance each other causing a lower $\not\!\!\!\:E_{\rm T}$ than in the
case of hadronic chargino decays, i.e.\ in the missing $E_{\rm T}$ plus
multijet channel.
After the above selection there is still a significant contribution from
mismeasured jets, and Standard Model processes. Kinematic cuts are used
to reduce their contribution:
$ \Delta \varphi ( \not\!\!\!\:E_{\rm T}, \mbox{jet}_{1} ) > 90^{\circ}$
and a combination of azimuthal opening angle of the leptons, $ \Delta
\varphi ( l_{1}, l_{2} )$ and transverse momentum of the dilepton system.

In the data two candidate events are found. According to a background
calculation $1.29 \pm 0.62 \mbox{(stat)} \pm 0.35 \mbox{(sys)}$ events
are expected in the sample from Standard Model processes. This agrees
within the uncertainty with our observation. Hence, we can only use
the result to set limits on gluino and squark production. To do so we
need to estimate the event expectation in different SUSY scenarios
and the systematic uncertainties associated with both background and
signal expectation. The SUSY expectation is estimated with the help of
Monte Carlo programs. The main systematic uncertainties in the signal
expectation are from the energy scale in the calorimeter (which effects
jet $E_{\rm T}$ and lepton isolation), about $\pm 5 \%$, from the
luminosity measurement and trigger efficiency, about $15 \%$, and from
the lepton identification, about $12 \%$.

\begin{figure}[tbp]
\psfig{figure=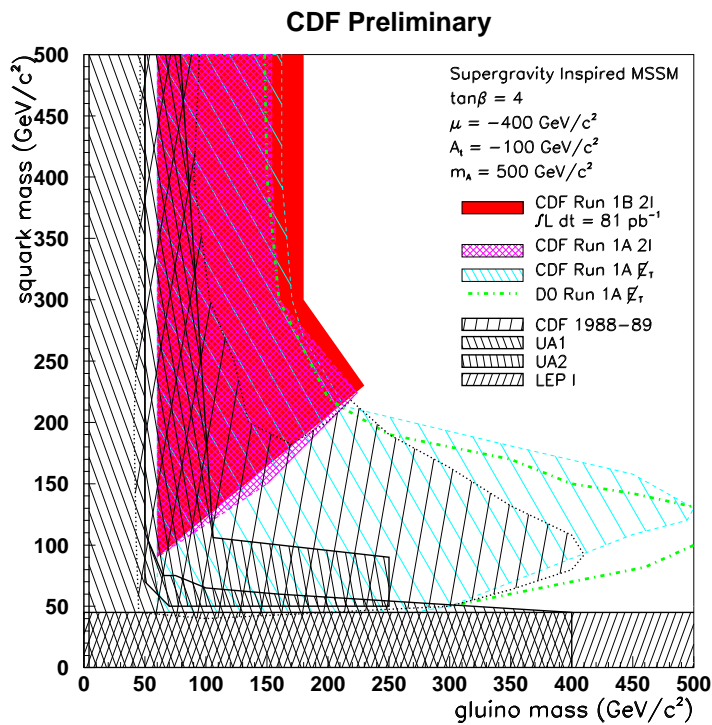,height=100mm}
\caption{Gluino and squark 95\% confidence level mass limits of the
CDF like-sign dilepton analysis.
\label{fig:cdf_dil_limit}}
\end{figure}

Figure~\ref{fig:cdf_dil_limit} shows the region in gluino--squark mass
excluded by the CDF dilepton analysis. The limits vary only slightly
as function of $\tan (\beta)$. For small $\left| \mu \right|$ values
the limit vanishes as the LSP becomes very Higgsino like. However,
this region is already excluded by $Z$ measurements at LEP.

\subsubsection{Single Lepton}
The single lepton channel is very interesting for future searches,
once gluinos and squarks are ``heavy enough''. The key variables in
the search are the missing $E_{\rm T}$ and the transverse mass of
the lepton---missing $E_{\rm T}$ system. The $W$ background has smaller
missing $E_{\rm T}$ and a transverse mass with a Jacobian peak at
the $W$ mass. Gluino and squark production, however, produces a much
harder missing $E_{\rm T}$ spectrum and a very wide transverse mass
distribution. Studies done at CDF show that this channel may be already
feasable at the Tevatron with Run I data.

\subsection{Charginos and Neutralinos}
\begin{figure}[tbp]
\hspace{20mm}
\psfig{figure=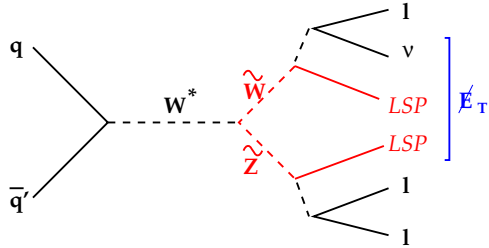,height=40mm}
\caption{Chargino--neutralino pair production (in $s$-channel) and decay
into final state with three charged leptons.
\label{fig:trilepton}}
\end{figure}

At the Tevatron, chargino--neutralino pair production occurres through
virtual $W$ in $s$-channel with some small negative interference from
the $t$-channel virtual squark diagrams (Figure~\ref{fig:trilepton}).
A wino-like chargino together with a very bino-like neutralino, i.e.\ the
second lightest neutralino, could have a significant production
cross-section up to a fraction of a ${\rm pb}$.

The chargino and neutralino mass ranges that are currently being probed
are close to the $W$ and $Z$ masses, i.e.\ rather light compared to the
gluino and squark masses of several hundred ${\rm GeV}/c^{2}$.
We can compare the situation of chargino--neutralino pair production to
$W$--$Z$ pair production. The initial state of the $s$-channel diagram
is identical. The chargino and neutralino now have spin $1/2$ versus
spin $1$ of the $W$ and $Z$. In the decay we have the additional LSP
for the chargino and neutralino. The LSP, much heavier than the leptons,
will take a significant fraction of the energy of the chargino and
neutralino. Thus, we expect a much softer lepton spectrum in the case
of chargino--neutralino production than in the case of $W$--$Z$
production.

Restricting the search to final states with three charged leptons makes
the signal very clean and almost background free. We are then searching
for events with one same generation $\ell ^{+} \ell ^{-}$ pair, and
additional lepton, and significant missing $E_{\rm T}$ from the neutrinos
and LSPs. The only hadronic energy in the event is from the spectator
partons or initial state radiation and gluon splitting. All three charged
leptons should be rather isolated from hadronic energy and the events
should have little jet activity if any.

The search for chargino--neutralino via trileptons has a small model
dependency. The chargino and neutralino mass and the neutralino branching
ratio into leptons are the main parameters. The LSP mass becomes important
only if it is very heavy. In minimal SUGRA models we expect the LSP mass
to be about half of the chargino mass and the neutralino and chargino to
be of similar mass. The leptonic branching ratio of the chargino is close
to the leptonic $W$ branching ratio if the decays through virtual $W$
dominate. Light sleptons can increase this branching ratio.

The CDF Run Ib analysis requires one central electron or muon of at least
$11 \, {\rm GeV}/c$ transverse momentum. With this lepton requirement the
event has a high efficiency for passing the single electron or muon
trigger of the experiment. Two more electrons with $E_{\rm T} > 5 \,
{\rm GeV}$ or muons
with $p_{\rm T} > 4 \, {\rm GeV}/c$ are required. For electrons this is
about the lowest $E_{\rm T}$ where electrons can be identified in the
detector with high efficiency. The first lepton has to satisfy stringent
identification requirements. For the second and third leptons a more loose
identification is allowed to retain high signal efficiency. The leptons
have to originate from a common vertex to reject events with leptons from
different interactions during the same beam-beam crossing and non-prompt
leptons. All three leptons must be isolated, where isolation is defined as
less than $2 \, {\rm GeV}$ total $\Sigma E_{\rm T}$ in the calorimeter inside
an $\eta$--$\varphi$ cone of radius $\Delta R = 0.4$. This is only half
the energy we allowed for isolation in the dilepton search above. In the
dilepton search we had a strong production process and expected large jet
activity. We now have a weak production and expect no jet activity and
thus, can tighten the isolation requirement without compromising signal
efficiency. There must be at least one $e^{+}e^{-}$ or $\mu ^{+} \mu ^{-}$
pair in the event from the neutralino decay. To remove background from
$b \overline{b}$ cascade decays we require the leptons to be seperated
by more than $\Delta R = 0.4$. High mass dileptons from the Drell-Yan
process favour a back-to-back topology in azimuthal angle. We reject any
event with opening angle between the two highest-$p_{\rm T}$ leptons
$\Delta \varphi > 170^{\circ}$. Events with an $e^{+} e^{-}$ or
$\mu ^{+} \mu ^{-}$ invariant mass of a known vector meson or the $Z$
resonance are removed. Finally a rather small missing $E_{\rm T}$ of
$15 \, {\rm GeV}$ is required.

\begin{table}[tbp]
\begin{center}
\caption{Events remaining after each cut in the CDF Run Ib trilepton
analysis. The chargino--neutralino expectation is for a mass of $70 \,
{\rm GeV}/c^{2}$, $\tan \beta = 2$, $\mu = - 400 \, {\rm GeV}/c^{2}$, and
$m_{\tilde{q}} = m_{\tilde{g}}$.
\label{tab:trilepton}}
\begin{tabular}{|l|r|c|c|}
                                                                 \hline
       & number of & expected SM  & possible                    \\
       & events    & contribution &
       $\tilde{\chi^{\pm}_{1}} \tilde{\chi^{0}_{2}}$ signal   \\ \hline
                                                                   \hline
 trilepton selection                 & $70$ &               &   \\ \hline
 vertex requirement                  & $59$ &               &   \\ \hline
 isolation $I < 4 \, {\rm GeV}$      & $23$ &               &   \\ \hline
 $e^{+}e^{-}$ or $\mu ^{+} \mu ^{-}$ & $23$ &               &   \\ \hline
 $\Delta R_{\ell \ell} > 0.4$        & $ 9$ &               &   \\ \hline
 $\Delta \varphi < 170^{\circ}$      & $ 8$ & $9.6 \pm 1.5$ & $6.2 \pm 0.6$
                                                                \\ \hline
 $J/\psi$, $\Upsilon$, $Z$ removal   & $ 6$ & $6.6 \pm 1.1$ & $5.5 \pm 0.5$
                                                                \\ \hline
 $\not\!\!\!\:E_{\rm T} > 15 \, {\rm GeV}$
                                     & $ 0$ & $1.0 \pm 0.2$ & $4.5 \pm 0.4$
                                                                \\ \hline
                                                                   \hline
 combining Run Ia and Run Ib:        & $ 0$ & $1.2 \pm 0.2$ & $5.5 \pm 0.4$
                                                                \\ \hline
\end{tabular}
\end{center}
\end{table}

No candidate events are observed in the $107 \, {\rm pb}^{-1}$ Run Ia
and Ib data of CDF. Chargino--neutralino expectation is simulated with
Monte Carlo methods, similar to the searches described above. The most
important issue in the chargino--neutralino search via trilepton events
is the lepton acceptance as it enters with the third power. The total
acceptance of the analysis described above varies between 2\% and
10\% for chargino and neutralino masses between $40$ and $90 \,
{\rm GeV}/c^{2}$.

\subsection{Third Generation Squarks}
The third generation of squarks and sleptons is rather special due to
the Yukawa couplings involved. As a result of the large top quark Yukawa
coupling the squarks of the third generation, sbottom and stop, could
be significantly lighter than the squarks in the first two generations.
This makes those two squarks rather special and warrants dedicated
search strategies. A single light squark would have a lower production
cross-section and thus a smaller signal and may escape searches assuming
five degenerate squarks. Two classes of searches open up:
\begin{itemize}
   \item Searches for direct production of $\tilde{t} \overline{\tilde{t}}$
         or $\tilde{b} \overline{\tilde{b}}$ pairs similar to the searches
         discussed above are possible.
   \item Searches for $\tilde{t}$ and $\tilde{b}$ in the decay products of
         other heavy particles like, for instance, top quarks are possible.
\end{itemize}
With the top quark being so very heavy, especially the second class of
searches are very appealing. We will discuss examples of both kinds in the
following sections.

The superpartners of the right-handed top quark and left-handed top quark
will not be mass eigenstates but mix. The result is a lighter stop,
$\tilde{t}_{1}$, and heavier stop state, $\tilde{t}_{2}$. Our searches are
then concentrating on the lighter mass state.
Similar mixing and mass splitting can occur for the sbottoms.

\subsubsection{Stop}
Stop production at the Tevatron is very similar to $t \overline{t}$
production: a strong interaction process with main contribution from
the valence quarks at high stop mass. However, for same masses the
$\tilde{t}_{1} \overline{\tilde{t}}_{1}$ production
cross-section~\cite{DESY_stop} is
about an order of magnitude smaller than $t \overline{t}$ production.
A factor of four comes from the spin difference (spin $0$ for stop
versus spin $1/2$ for top) and a factor two due to the fact that we
are only searching for the lighter stop state.

Depending on the stop, chargino, and slepton masses we can have the
following decay scenarios:
\begin{itemize}
   \item $\tilde{t}_{1} \longrightarrow \tilde{\chi}^{+}_{1} + b$
         followed by the decay of the chargino, e.g.\
         $\tilde{\chi}^{+}_{1} \longrightarrow \ell ^{+} + \nu +
         \mbox{LSP}$;
   \item in the case of a heavy chargino, i.e.\ $m_{\tilde{t}_{1}} <
         m_{\tilde{\chi}^{+}_{1}} + m_{b}$:
         \begin{eqnarray*}
         \tilde{t}_{1} & \longrightarrow & \tilde{\ell}^{+} + \nu + b \\
                       & \longrightarrow & \ell ^{+} + \tilde{\nu} + b;
         \end{eqnarray*}
   \item if the sleptons are heavy too, i.e.\ $m_{\tilde{t}_{1}} <
         m_{\tilde{\ell}} + m_{b}$, and $m_{\tilde{t}_{1}} <
         m_{\tilde{\nu}} + m_{\ell} + m_{b}$, then:
         \[ \tilde{t}_{1} \longrightarrow c + \mbox{LSP} \]
\end{itemize}
The first two cases can result in final states with two leptons, two
$b$-jets, and missing $E_{\rm T}$ from the LSPs and neutrinos. In the
case of the third scenario we get a very different signature: two
charm jets and missing $E_{\rm T}$.

A dilepton based search and a single lepton plus B-tag
search~\cite{Baer_stop} address the first two decay scenarios. Both
analysis are very challenging. The small
$\tilde{t} \overline{\tilde{t}}$ production cross-section, the soft
lepton momentum spectrum, and the soft missing $E_{\rm T}$ (due to the
four sources balancing each other) require sophisticated analyses to
identify any stop contribution hiding in the much larger Standard Model 
background from $t \overline{t}$, $b \overline{b}$ production, and the
Drell-Yan process.

In the dilepton analysis the leptons are of opposite electric charge, i.e.\
a rather common signature. The leptons are rather soft because the stop
masses that are currently being probed are below $100 \, {\rm GeV}/c^{2}$
and because of the LSP in the decay. The minimum lepton momenta in the
analyses are driven by trigger requirements and identification efficiency:
an electron or muon around $p_{\rm T} > 10$ to $15 \, {\rm GeV}/c$ and
a second lepton with $p_{\rm T}$ above $3$ to $8 \, {\rm GeV}/c$ are
required. We only expect a soft missing $E_{\rm T}$ in the signal events.
A requirement of around $15 \, {\rm GeV}$ can be made without
compromising signal efficiency. The $\not\!\!\!\:E_{\rm T}$ is
important in rejecting events from the Drell-Yan process. However, a
$15 \, {\rm GeV}$ cut is rather soft and detector effects can produce
such a rather small energy imbalance. The two $b$-jets in the events are
also of rather low energy.
The key variables in the analysis are the isolation of the leptons and
the ``bigness''. The leptons are coming from chargino or slepton decays and
are isolated from hadronic energy. On the other hand, leptons from the
semileptonic decay of bottom quarks are surrounded by hadronic energy.
The bigness is defined as scalar sum of lepton transverse momenta and
missing $E_{\rm T}$. For $t \overline{t}$ events the variable has large
values. For $b \overline{b}$ events the value is rather small. Our
$\tilde{t} \overline{\tilde{t}}$ events sit in the middle with modest
bigness values. The D\O\ experiment has reported results from a
dielectron search~\cite{D0_r1b_stop_diele}. The CDF analysis is still
in progress.

The single lepton plus B-tag analysis relies on only one leptonic chargino
decay. To improve the signature, one jet is required to originate
from a vertex with a measurable distance to the primary vertex of the
collision, consistent with the B hadron lifetime. The silicon vertex
detector in the CDF experiment allows such a measurement, i.e.\ B-tagging.
For high-$E_{\rm T}$ $b$-jets the tagging efficiency reaches up to 
about $30\%$. The CDF analysis started from the top dataset which had
a lepton $p_{\rm T}$ requirement of $20 \, {\rm GeV}/c$. After missing
$E_{\rm T}$ and two jets requirements (with one jet being B-tagged) a
likelihood variable is constructed using the transverse mass of the
lepton and $\not\!\!\!\:E_{\rm T}$ and the opening angle in the transverse
plane between lepton and second jet, $\Delta \varphi$.
Fitting the likelihood distribution of the data with the expected Standard
Model contribution and a stop expectation shows no significant stop
contribution in the data.
The lower $95\%$ confidence level cross-section limit obtained from the
fit is above $\tilde{t} \overline{\tilde{t}}$ expectation and does not
translate into a mass limit.
The low sensitivity of the analysis is due to the high-$p_{\rm T}$
requirement of the lepton. Lowering the transverse momentum requirement
of the first lepton is currently in progress and should result in a
significant increase in sensitivity for this search.

\begin{figure}[tbp]
\psfig{figure=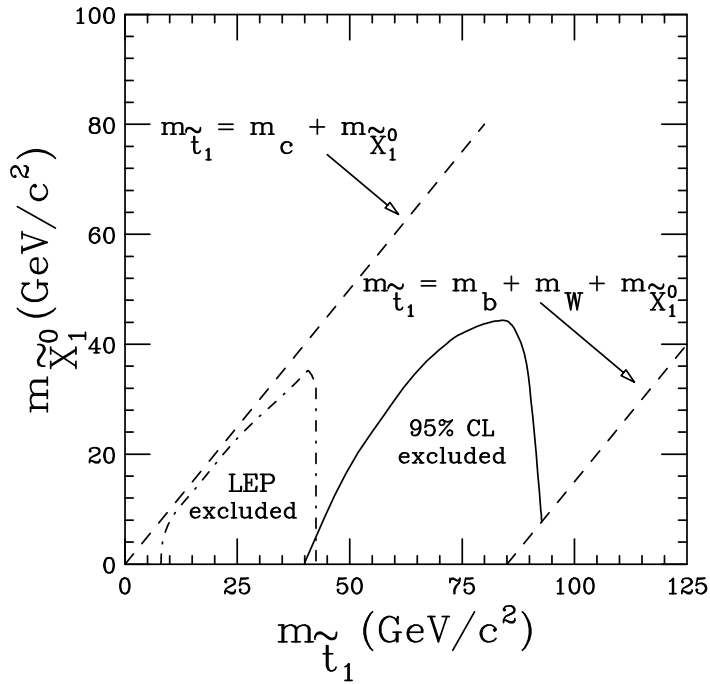,height=100mm}
\caption{The $95\%$ confidence level exclusion contour of the D\O\ missing
$E_{\rm T}$ plus two jets stop search.
\label{fig:d0_stop}}
\end{figure}

Stop pair production with the third decay scenario,
$\tilde{t}_{1} \longrightarrow c + \mbox{LSP}$ is addressed through
missing $E_{\rm T}$ plus two jets based searches. The D\O\  Run Ia
analysis is similar to the three and four jet analysis described
above. Two jets with transverse energy of at least $30 \, {\rm GeV}$
and a $\not\!\!\!\:E_{\rm T}$ larger than $40 \, {\rm GeV}$ are required.
Mismeasured QCD dijet events dominate the event sample. Hard cuts on
the opening angle between $\not\!\!\!\:E_{\rm T}$ and jet are needed
to eliminate this background: $\Delta \varphi ( \not\!\!\!\:E_{\rm T},
\mbox{jet} ) > 10^{\circ}$, $\Delta \varphi ( \not\!\!\!\:E_{\rm T},
\mbox{jet}_{1} ) < 125^{\circ}$, and $90^{\circ} < \Delta \varphi (
\mbox{jet}_{1}, \mbox{jet}_{2} ) < 165^{\circ}$.
Events with identified electrons or muons are vetoed as in the three
and four jet analysis. Two events are found in $7.4 \, {\rm pb}^{-1}$
of data. From Standard Model sources $2.86 \pm 0.93$ events are expected,
mainly $W$ (plus jets) production.
The sensitivity to $\tilde{t} \overline{\tilde{t}}$ contribution is 
calculated as function of stop and LSP mass (assuming $100\%$ branching
ratio of $\tilde{t}_{1} \longrightarrow c + \mbox{LSP}$).
Figure~\ref{fig:d0_stop} shows the $95\%$ confidence limit of the D\O\
analysis.

With the top quark being very heavy and the top squark being potentially
rather light, there is a possibility of the top quark to decay not only
into $W$ and $b$ but also into $\tilde{t}$ and LSP. The CDF and D\O\ top
analyses are based on the assumption of a Standard Model top decay. This
raises the question if there are any $t \overline{t}$ events with
$t \longrightarrow \tilde{t} + \mbox{LSP}$ decays at the Tevatron.
If none are observed, how much branching ratio could hide from us?

\begin{figure}[tbp]
\hspace{15mm}
\psfig{figure=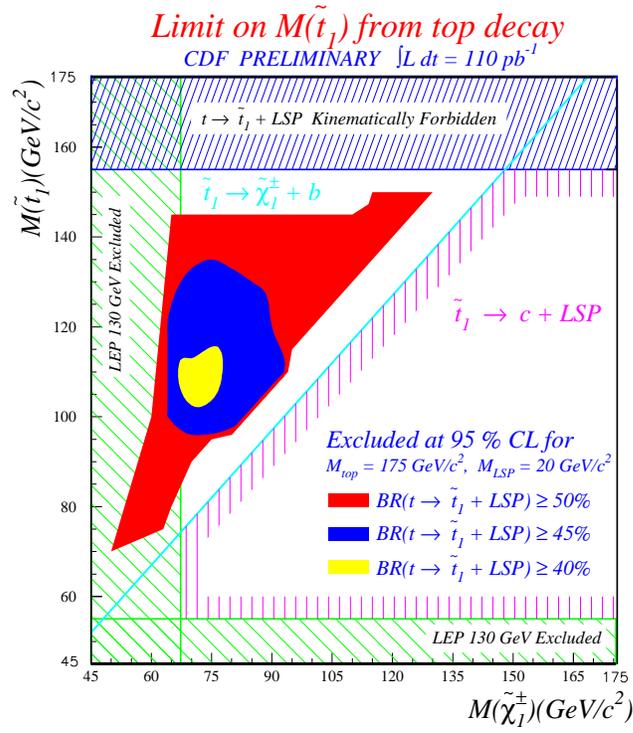,height=100mm}
\caption{The $95\%$ confidence level excluded region of stop--chargino
mass as function of $t \longrightarrow \tilde{t} + \mbox{LSP}$ branching
ratio for a LSP mass or $20 \,{\rm GeV}/c^{2}$.
\label{fig:cdf_topstop}}
\end{figure}

An analysis done by the CDF collaboration addresses the case where the
chargino is lighter than the stop, i.e.\ the decay $\tilde{t}_{1}
\longrightarrow \tilde{\chi}^{+}_{1} + b$ is kinematically allowed.
The analysis is a combination of the standard single lepton plus
B-tag~\cite{cdf_top} and single lepton kinematic
analysis~\cite{CDF_top_kinematic}.
The analysis targets events where one of the top quarks decays into
$W + b$ and the other top quark into supersymmetric particles, $\tilde{t}
+ \mbox{LSP}$. As long as neither branching ratio is very large/small
most $t \overline{t}$ events will have such a decay combination. In the
case of a leptonic $W$ decay, the top quark with Standard Model decay
yields an energetic lepton and thus satisfies all trigger and selection
requirements as in the standard single lepton top analysis. The second
top quark is then used as a probe. Compared to Standard Model decay we
now have two LSPs and an additional step in the decay:
\[ \begin{array}{llcll}
 t \longrightarrow & \tilde{t} + \mbox{LSP} &
 \; \; \; \; \; \; \; \; \; \; \; \; \; \; \; \; \; \; \; \; \; \; &
 t \longrightarrow & W + b \\
                   & \, \hookrightarrow \tilde{\chi}^{+} + b & &
                   & \; \hookrightarrow q \overline{q}' \\
             & \; \; \; \; \; \; \hookrightarrow q \overline{q}' + \mbox{LSP}
\end{array} \]
The quark jets will have a much softer energy spectrum in the case of a
SUSY top decay. The $E_{\rm T}$ spectrum of the third and fourth highest
$E_{\rm T}$ jet shows a significant difference between events where both
top quarks decayed into $W + b$ and events where one top quark decayed
into $\tilde{t} + \mbox{LSP}$. The $E_{\rm T}$ information of the two jets 
is combined in a likelihood variable for best seperation. All observed
events fall in the $W + b$ like region. The contribution to the $\tilde{t}
+ \mbox{LSP}$ like region is evaluated as function of stop, chargino, and
LSP masses. Figure~\ref{fig:cdf_topstop} shows the branching ratio of
$t \longrightarrow \tilde{t} + \mbox{LSP}$ excluded at $95\%$ confidence
level by this analysis.

\subsubsection{Sbottom}
Another example of a search for superpartners in the decay of other
particles, this time gluinos, is the search for bottom squarks. There
are two advantages over a search for direct sbottom pair production:
The cross-section for gluino pair production is significantly larger
than for direct $\tilde{b}_{1} \overline{\tilde{b}}_{1}$ production
and second the signature is much richer.

In the dilepton based gluino search we assumed a gluino decay of
$\tilde{g} \longrightarrow q \overline{q}' + \tilde{\chi}^{\pm}$. If
the sbottom is significantly lighter than the other squarks, the two
body decay $\tilde{g} \longrightarrow \tilde{b} \overline{b}$ could
be kinematically allowed. A light stop, on the other hand, will not
work due to the heavy top quark, i.e.\ the stop would need to be $175
\, {\rm GeV}/c^{2}$ lighter than the gluino.
The sbottom itself would decay into a bottom quark and the LSP. The
signature is then four $b$-jets and significant missing $E_{\rm T}$.
$b \overline{b}$ production with hard gluon radiation and splitting
into $b \overline{b}$ is the main Standard Model process yielding
four $b$-jets. The $\not\!\!\!\:E_{\rm T}$ would have to come from
jet mismeasurements.
The $b$-jets can be identified through semileptonic decays or via the
secondary vertex method described above. Identifying two or three of
the four $b$-jets in the event yields a high signal efficiency and
allows good background rejection. An analysis at CDF based on the
secondary vertex method is in progress.

\subsection{R-Parity violating SUSY}
Searches for R-parity violating supersymmetry are also possible at
the Fermilab Tevatron~\cite{Baer_Rparity}. The excess of events with
high $Q^{2}$ at HERA~\cite{HERA_high_q2} has triggered several new
analyses in both CDF and D\O . First preliminary results have been
reported~\cite{CDF_Rparity}. Moderate R-parity violating terms in the
Lagrangian impact mainly the decay of the superpartners. The terms
correspond to lepton number or baryon number violating processes.

Lepton number violating terms can result in a significant increase
in lepton production. The three-lepton terms, $\lambda _{ijk}$ can
cause decay of the LSP into three leptons. If this decay occurres
inside the detector all superpartner production processes would end
in multilepton final states. (If the decay occurres outside the
detector we have the standard missing $E_{\rm T}$ signatures.) The
lepton momenta will be in a similar range as in the case of
chargino--neutralino
production. The leptons will be isolated from hadronic energy.
However, the event may have substantial jet activity from the gluino
or squark cascade.

The one-lepton two-quark terms, $\lambda '_{ijk}$, can cause the
decay of squarks into quark plus lepton. They can also lead to
the decay of the LSP into a quark, antiquark, and lepton. In the
case of squark decays into quark plus lepton the lepton would
be rather energetic. In the case of LSP decays we expect more
moderate lepton momenta. The $\lambda '_{121}$ and $\lambda '_{131}$
terms are the ones of interest in the interpretation of the HERA
event excess as R-parity violating supersymmetry.

The three-quark terms, $\lambda ''_{ijk}$ correspond to baryon number
violating processes. The terms resulting in bottom quark final states
might be the only ones accessible at the Fermilab Tevatron.

\subsection{Photon Based Searches}
\begin{figure}[tb]
\psfig{figure=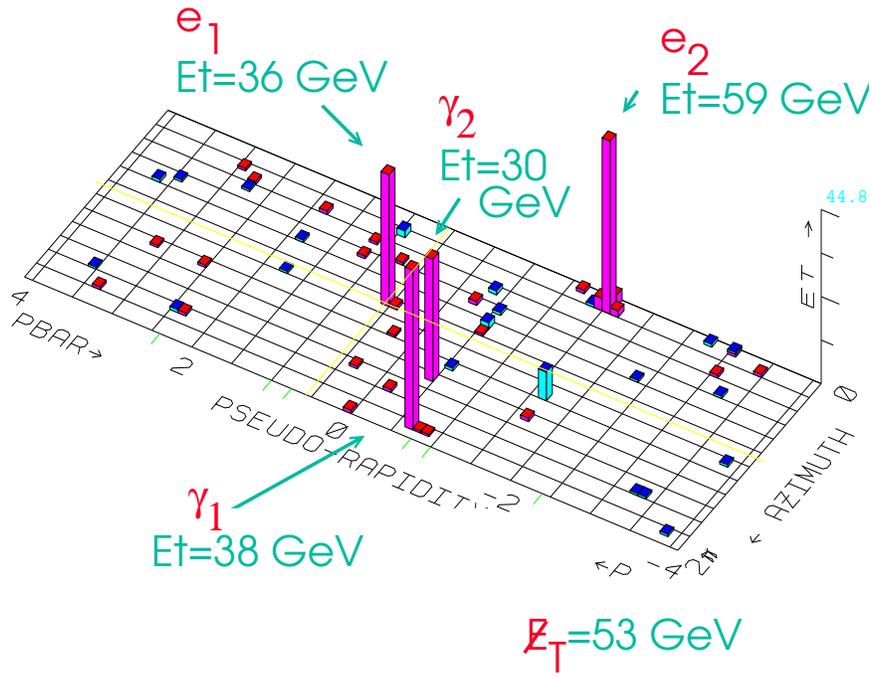,height=95mm}
\caption{LEGO plot of the famous CDF dielectron--diphoton--%
$\not\!\!\!\:E_{\rm T}$ candidate event.
\label{fig:cdf_eeggmet}}
\end{figure}

Photon based SUSY searches at hadron colliders are actually not a new
idea~\cite{CDF_rad_neutralino}. However, after an event from the CDF
experiment with two electron candidates, two photon candidates, and
an energy imbalance became public, interest increased
strongly~\cite{eeggmet_event}. A LEGO
plot of the event is shown in figure~\ref{fig:cdf_eeggmet}. The plot
shows the energy deposition in the ``un-rolled'' detector, i.e.\ in
$\eta - \varphi$ space. Apart from the four electromagnetic objects,
the event is very quiet. The four electromagnetic clusters, however,
are each very energetic, between $30$ and $60 \, {\rm GeV}$. Three of
the four objects are in the central region of the detector, the second
electron candidate is in the end-cap region of the detector and has
thus limited tracking information. The initial classification of this
object as electron was based on hits found in the vertex chamber in the
direction of the electromagnetic cluster. More detailed studies during
1996 showed that those hits are not aligned with the cluster. In the
case of this cluster originating from a prompt electron, there should be also
a track (or at least hits) in the silicon vertex detector. However, the
SVX has no track or hits in the direction of this cluster either.

What is the origin of the electromagnetic cluster and this event? The
electromagnetic cluster is unlikely to be from a prompt electron. It
is also not likely to be the result of a jet fluctuation or tau decay.
It could be a third photon. All those interpretations are possible
and we can calculate their probability for this one event (inside the
Standard Model).
As discussed earlier, an event-by-event analysis is almost impossible
at hadron colliders (with the exception of very striking signatures).
Four energetic electromagnetic objects in an event are very rare. There
is no Standard Model process that produces such events with significant
rate. One event in $100 \, {\rm pb}^{-1}$ could stand for one event
every $100 \, {\rm pb}^{-1}$ but also for one event in $10 \,
{\rm fb}^{-1}$ or $100 \, {\rm fb}^{-1}$, i.e.\ us just getting the one
event ``early''.

\begin{figure}[tbp]
\hspace{14mm}
\psfig{figure=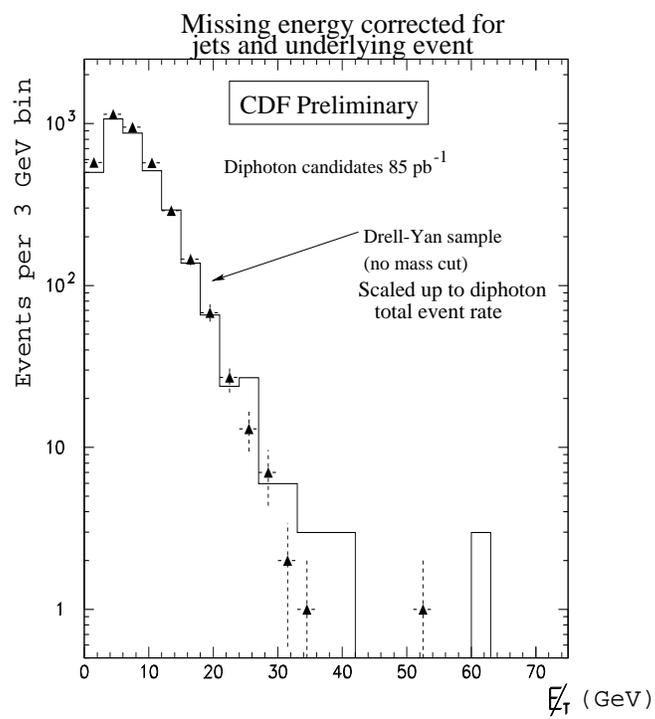,height=100mm}
\caption{The CDF missing $E_{\rm T}$ spectrum of diphoton ($E_{\rm T} >
12 \, {\rm GeV}$) candidate events.
\label{fig:cdf_ggmet}}
\end{figure}

\begin{figure}[tbp]
\hspace{6mm}
\psfig{figure=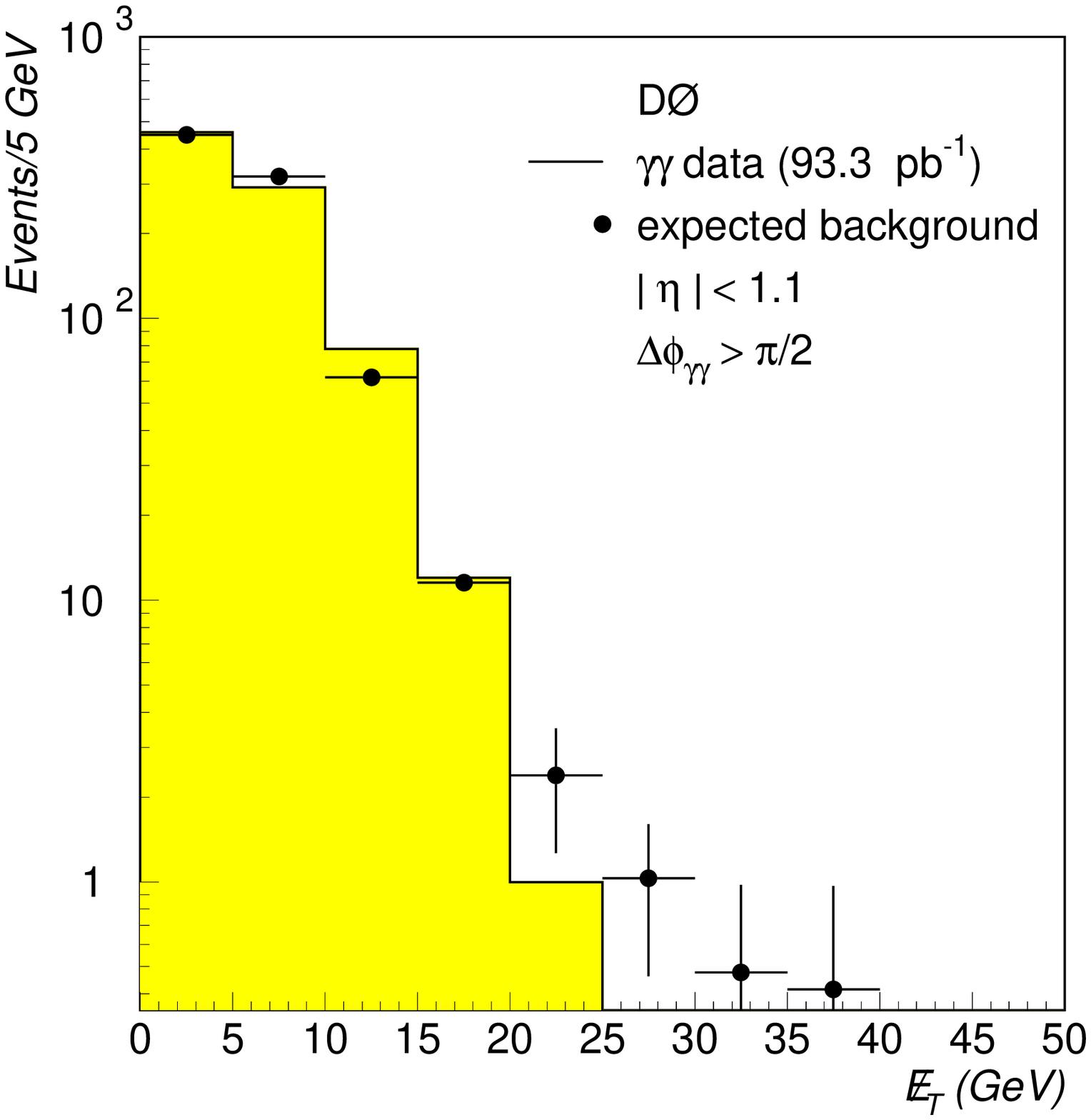,height=100mm}
\caption{The D\O\ missing $E_{\rm T}$ spectrum of diphoton events.
\label{fig:d0_ggmet}}
\end{figure}

For interpretations here, let's consider the electromagnetic object in
the end-cap region to be a prompt electron (since this was done in most
of the discussions of last year).
In the Standard Model $W W \gamma \gamma$ production could yield the
observed signature. CDF and D\O\ have observed first diboson events,
i.e.\ $W W$ and $W \gamma$.
A simple calculation shows that less than $0.00008$ events with
$e e \gamma \gamma \not\!\!\!\:E_{\rm T}$ are expected from this process.
If there were an anomoulous $W W \gamma \gamma$ production, we would
not only expect $e e \gamma \gamma \not\!\!\!\:E_{\rm T}$ events but also
$\ell \, \mbox{jet} \, \mbox{jet} \, \gamma \gamma \not\!\!\!\:E_{\rm T}$
events,
i.e.\ where one of the $W$s decayed hadronically (or $\gamma \gamma$ plus
jets events where both $W$s decayed hadronically).
We expect to find more than an order of magnitude more single
lepton--diphoton--$\not\!\!\!\:E_{\rm T}$ events than
dilepton--diphoton--$\not\!\!\!\:E_{\rm T}$ events if the observed event
is representative of $W W \gamma \gamma$ production. Both the CDF and D\O\
experiment have searched for additional diphoton--$\not\!\!\!\:E_{\rm T}$
events in the Run I data~\cite{D0_diphoton,CDF_diphoton}. None were found.
Figures~\ref{fig:cdf_ggmet}
and \ref{fig:d0_ggmet} show the missing $E_{\rm T}$ spectrum of diphoton
events from the two experiments.

There are (mainly) two SUSY hypotheses for the event. The first is based
on the assumption of a light gravitino and it being the lightest
supersymmetric particle. The second postulates a light Higgsino with it
being the LSP. In the light gravitino scenario the event could originate
from either selectron pair production or wino pair production (the state
is now a very pure superpartner of the $W$):
\[ \begin{array}{lclcl}
\tilde{e} & \longrightarrow & e & + & \tilde{B} \\
          &                 &   &   & \, \hookrightarrow \tilde{G} \gamma
\end{array} \]
\[ \begin{array}{lcl}
\tilde{W} & \longrightarrow & W + \tilde{B}
                                    \; \rightarrow \tilde{G} \gamma \\
          &                 & \; \hookrightarrow \ell \nu
\end{array} \]
The selectron or wino decay proceeds through a neutralino that is almost
pure bino into the gravitino and photon. In this scenario, the mass of the
gravitino would be around $m_{\tilde{G}} \approx 1 \, {\rm keV}/c^{2}$.
If the observed event is not due to statistical fluctuation but typical
of such a light gravitino scenario, then many more SUSY events, decaying
into final states with photons and gravitinos should be observed.

In the light Higgsino scenario the lightest supersymmetric particle is
almost a pure Higgsino state. The selectron cascades through an almost
pure gaugino state into the Higgsino and photon:
\[ \begin{array}{lclcl}
\tilde{e} & \longrightarrow & e & + & \tilde{Z}_{2} \\
          &                 &   &   & \, \hookrightarrow \tilde{H}_{1} \gamma
\end{array} \]
The scenario requires a gaugino mass of around $m_{\tilde{Z}_{2}} \approx
60 - 90 \, {\rm GeV}/c^{2}$ and a Higgsino mass of $m_{\tilde{H}_{1}}
\approx 35 - 55 \, {\rm GeV}/c^{2}$ to explain the event observed by CDF.
In this interpretation gaugino mass unification is no longer possible.

Both experiments are studying their sensitivity to various superpartner
production in the above two SUSY scenarios. The D\O\ experiment has
reported a seach for chargino and neutralino production in SUSY models
with a light gravitino~\cite{D0_light_grav}.

\section{Future at the Tevatron}
The Fermilab Tevatron is currently being upgraded to higher luminosity.
The Main Ring is being replaced by the Main Injector, a $150 \, {\rm GeV}$
proton synchrotron. The Main Injector will provide increased protons to
both antiproton production and the Tevatron. With an increased luminosity
from the $5 \cdot 10^{30}\, {\rm cm}^{-2} {\rm s}^{-1}$ of Run I to
$10^{32} \, {\rm cm}^{-2} {\rm s}^{-1}$ the
number of multiple collisions per beam-beam crossing would rise. To
counteract this, protons and antiprotons will be distributed over $106$
bunches instead of the current $6$ bunches. This will result in a much
shorter bunch spacing, $132 \, {\rm n} \mbox{sec}$ instead of the current
$3.5 \, \mu \mbox{sec}$.

The detectors cannot handle such a short bunch spacing and the higher
luminosity in their Run I configuration. Both CDF and D\O\ experiment
are being upgraded to cope with the changed conditions. Both detectors
require new front-end electronics to handle the shorter bunch spacing.
The tracking system of the CDF detector cannot withstand the higher
luminosity and is being replaced. The new tracking system of the D\O\
detector will be embeded in a magnetic field of $2 \, \mbox{Tesla}$.
For both experiments the muon detection will be improved: for D\O\ with
additional muon trigger detectors and for CDF with additional chambers
at larger pseudorapidity.

The next collider run is scheduled to start in the beginning of 2000.
The goal is to collect $2 \, {\rm fb}^{-1}$ of proton--antiproton
interactions with each experiment. In the search for supersymmetry
the data will allow us to significantly extend the reach of the experiments.
For charginos and neutralinos sensitivity will be increased up to
almost $200 \, {\rm GeV}/c^{2}$ and gluino and squark mass limits up
to about $350 \, {\rm GeV}/c^{2}$ are expected. Chances are good to
see the next symmetry of nature very early in the $21^{\rm st}$ century.

\section*{Acknowledgments}
It was a pleasure to lecture at the Theoretical Advanced Study Institute
in Boulder. I like to thank Jonathan Bagger and K.T.\ Mahanthappa for
the excellent organization and hospitality and the TASI students for
their interest in the subject. I am grateful to Joel Butler for close
reading of the manuscript and his comments.


\begin{thebibliography}{99}
\bibitem{coordinates}
  In the CDF and D\O\ coordinate systems, $\theta$ is the polar angle
  with respect to the proton beam direction ($z$), $\varphi$ is the
  azimuthal angle.
  The pseudorapidity $\eta$ is defined as $- \ln \tan (\theta / 2)$. The
  transverse momentum of a particle is then $p_{\rm T} = p \sin \theta $.
\end{thebibliography}
\end{document}